\documentclass[11pt,a4paper]{article}

\newcommand{\vev}[1]{\langle #1\rangle}
\newcommand{\wh}{\widehat}

\usepackage{epsfig,graphicx}
\usepackage{amsmath,amsfonts,amssymb}
\usepackage{citesort}
\usepackage{amsmath}
\usepackage[width=1.15\textwidth, small]{caption}

\newcommand{\bmat}{\left(\begin{array}}
\newcommand{\emat}{\end{array}\right)}

\def\lsim{\raise0.3ex\hbox{$\;<$\kern-0.75em\raise-1.1ex\hbox{$\sim\;$}}}
\def\gsim{\raise0.3ex\hbox{$\;>$\kern-0.75em\raise-1.1ex\hbox{$\sim\;$}}}

\def\yzero{\smash{\hbox{$y\kern-4pt\raise1pt\hbox{${}^\circ$}$}}}

\def\b{\beta}

\def\s2{\frac{1}{\sqrt2}}
\def\wh{\widehat}

\def\beq{\begin{equation}}
\def\eeq{\end{equation}}
\def\beqa{\begin{eqnarray}}
\def\eeqa{\end{eqnarray}}

\def\IF{\relax{\rm I\kern-.18em F}}
\def\II{\relax{\rm I\kern-.18em I}}
\def\IP{\relax{\rm I\kern-.18em P}}
\def\IC{\relax\hbox{\kern.25em$\inbar\kern-.3em{\rm C}$}}
\def\IR{\relax{\rm I\kern-.18em R}}

\def\Dsl{\,\raise.15ex\hbox{/}\mkern-13.5mu D} 
\def\IZ{Z\kern-.4em  Z}

\def\bmat{\left(\begin{array}}

\def\emat{\end{array}\right)}

\def    \part          {\partial}
\def    \be            {\begin{equation}}
\def    \ee            {\end{equation}}
\def    \bea           {\begin{eqnarray}}
\def    \eea           {\end{eqnarray}}

\begin{document}

\begin{flushright}
FTUAM 07/13\\
IFT-UAM/CSIC-07-44\\
LPT-ORSAY-07-72\\
arXiv:0710.3672 [hep-ph]\\
\vspace*{3mm}
{\today}
\end{flushright}

\vspace*{5mm}
\begin{center}
{\Large \textbf{Lepton masses and mixings in orbifold models \\
with three Higgs families}}

\vspace{0.5cm} 
{\large N.~Escudero$\,^{a,b}$, C.~Mu\~noz$^{\,a,b}$ and
  A.~M.~Teixeira$\,^c$}\\
[0.2cm]
{$^a$\textit{Departamento de F\'{\i}sica Te\'{o}rica C-XI, Universidad
    Aut\'{o}noma de Madrid,\\ 
Cantoblanco, E-28049 Madrid, Spain}}\\
\vspace*{0.2cm} 
{$^b$\textit{Instituto de F\'{\i}sica Te\'{o}rica UAM/CSIC, Universidad
    Aut\'{o}noma de Madrid,\\ 
Cantoblanco, E-28049 Madrid, Spain}}\\[0pt]
\vspace*{0.2cm} 
{$^c$\textit{Laboratoire de Physique Th\'eorique, UMR 8627, Universit\'e de
    Paris-Sud XI, B\^atiment 201, \\ 
F-91405 Orsay Cedex, France}}\\[0pt]

\vspace*{0.3cm} 
\end{center}
%
%
\abstract
We analyse the phenomenological viability of heterotic $Z_3$ 
orbifolds with two Wilson lines, which naturally predict three 
supersymmetric families of matter and Higgs fields. Given that these
models can accommodate realistic 
scenarios for the quark sector avoiding potentially
dangerous flavour-changing neutral currents, 
we now address the
leptonic sector, finding that viable orbifold configurations
can in principle be obtained. 
In particular,
it is possible to accomodate present data on charged lepton
masses, while avoiding conflict with lepton flavour-violating decays.
Concerning the generation of neutrino masses and mixings, we find that
$Z_3$ orbifolds offer several interesting possibilities. 

%
%
\section{Introduction}

At present, string theory is the only candidate to unify all 
known interactions (strong, electroweak and gravitational) in 
a consistent way. In addition to necessarily containing 
the standard model (SM) as its low-energy limit, string theory
is expected to address some of its puzzles, namely
the nature of masses, mixings and number of families.

From observation, we have firm evidence that Nature contains three
families of quarks and leptons, with peculiar mass hierarchies. 
Moreover, the flavour structure
in both quark and lepton sectors is far from trivial, as exhibited by
the current bounds on the quark~\cite{pdg2004} and
lepton~\cite{GonzalezGarcia:2007ib,Maltoni:2004ei,Strumia:2005tc,Fogli:2005cq}  
mixing matrices. At present, we are still lacking two ingredients that  
are instrumental in understanding flavour dynamics: the discovery of the
Higgs boson (which would thus confirm the mechanism of mass generation), 
and the precise knowledge of how fermions and Higgs scalars interact, that
is, the Yukawa couplings of the fundamental theory. In this sense, a
theory that aims at successfully explaining the observed fermion
spectra, must necessarily be predictive regarding the Yukawa
couplings. 

Within the context of string theory, a natural and aesthetic solution
may arise from $Z_3$ orbifold compactifications. In fact, 
a very interesting way to obtain a
four dimensional effective theory is the compactification of the $E_8
\times E_8$ heterotic string~\cite{Gross:1984dd} on six-dimensional
orbifolds~\cite{Dixon:1986jc}, and this has proved to be a very successful
attempt at finding the superstring standard 
model~\cite{Ibanez:1986tp,Ibanez:1987sn,Bailin:1987xm,Ibanez:1987pj,Casas:1987us,Font:1988tp,Kim:1988dd,Casas:1988se,Casas:1988hb,Font:1988mm,Casas:1988vk,Casas:1988wy,Font:1989aj,Casas:1989wu,Katsuki:1989bf,Kim:1992en,Aldazabal:1995cf,Giedt2,Munoz:2001yj,Raby,Giedt:2005vx,Kobayashi:2005vb,Lebedev,Kim,Lebedev2,Kim25,hum,Lebedev3}.
As it was shown in~\cite{Ibanez:1987sn,Ibanez:1987pj}, the
use of two Wilson
lines~\cite{Dixon:1986jc,Ibanez:1986tp} on the torus defining
a symmetric $Z_3$ orbifold can give rise to supersymmetric (SUSY) models with
$SU(3)\times SU(2)\times U(1)^n$ gauge group and three families of
chiral particles with the correct $SU(3)\times SU(2)$ quantum numbers.
These models present very attractive features from a phenomenological
point of view. 
One of the $U(1)$s of the extended gauge group is in general
anomalous, and it can induce a Fayet-Iliopoulos (FI) 
$D$-term~\cite{Witten:1984dg,Dine:1987xk,Atick:1987gy,Dine:1987gj} that
would break SUSY at very high energies (FI scale $\sim
\mathcal{O}(10^{16-17}$ GeV)). To preserve SUSY, some fields
will develop a vacuum 
expectation value (VEV) to cancel the undesirable $D$-term. The FI
mechanism allows to break the
gauge group down to $SU(3)_c\times SU(2)_L\times U(1)_Y$, as shown in 
Refs.~\cite{Casas:1988hb,Font:1988mm} and~\cite{Casas:1987us}.

Orbifold compactifications offer some remarkable properties in
relation to the flavour problem. In particular, 
they provide a geo\-me\-tric mechanism to
generate the mass hierarchy for quarks and 
leptons~\cite{Hamidi:1986vh,Dixon:1986qv,Ibanez:1986ka,Casas:1989qx,Casas:1992zt} through renormalisable Yukawa couplings. 
$Z_n$ orbifolds have twisted fields which are attached to the 
orbifold fixed points. Fields at different fixed points may
communicate with each other only by world sheet instantons. The
resulting renormalisable Yukawa couplings can be explicitly 
computed~\cite{Hamidi:1986vh,Dixon:1986qv,Casas:1990hi,Burwick:1990tu,Kobayashi:1991rp,Casas:1991ac,Abel:2002ih,Ko:2004ic,Ko:2005sh}
and receive exponential suppression factors that depend on the 
distance between the fixed points to which the relevant fields are 
attached. These distances can be varied by giving different VEVs
to the $T$-moduli associated with the size 
and the shape of the orbifold. 

Models with purely renormalisable Yukawa couplings, and which are
a priori successful from a phenomenological point of view can be obtained if
one relaxes the requirement of a minimal SUSY matter content (with
just two Higgs doublets)~\cite{Abel:2002ih}. 
In fact, $Z_3$ orbifolds with two Wilson lines
naturally contain three families of everything, including Higgses, and 
allow for realistic fermion masses and mixings, entirely at the renormalisable
level. In this case the low-energy theory corresponds to the minimal
supersymmetric standard model (MSSM) with three Higgs families. 

Whether or not such values are indeed viable from a phenomenological
point of view is a question that deserves careful consideration. 
Not only should these models correctly reproduce quark
and lepton masses and mixings, but they should also comply with the
current bounds on flavour-changing neutral currents
(FCNCs). Indeed, having three Higgs families, with non-trivial
couplings to matter potentially gives rise to FCNCs at the 
tree-level~\cite{Georgi:1978ri,MWL-80,Shanker-81,EMT-1_06}.
In previous studies~\cite{EMT-2_06}, we have addressed the 
phenomenology of the 
Higgs and quark sectors of $Z_3$ models with two Wilson lines. We have
verified that quark masses and mixings could be reproduced, and
that FCNCs in the neutral $K$-, $B$- and $D$-sectors could be avoided
by a fairly light Higgs spectrum.
It is worth recalling that, after fitting the quark data, the free
parameters defining the orbifold geometry are already very
constrained.

In this work, we complete our previous analyses, by
investigating whether or not the $Z_3$ orbifolds can also succeed in
accommodating present data on charged lepton masses, while avoiding
conflict with lepton flavour violating (LFV) tree-level processes,
such as three-body decays. 
Finally, and most interesting, is the question of complying with
experimental data 
on neutrino mass squared differences and mixing angles. This
analysis is particularly challenging since, as we will discuss,
orbifolds offer a variety of possibilities to account for the smallness
of neutrino masses.

Let us recall that
the experimental observation of neutrino oscillations has led to
extend the SM in order to accommodate non-vanishing neutrino
masses. In the absence of a predictive theory for the Yukawa
couplings, it is only common to argue that purely Dirac neutrinos 
pose a naturalness problem, in the sense that the associated couplings
are extremely tiny. Moreover, and contrary to what is observed in the
quark sector, the leptonic mixing, parameterised by the
Maki-Nagakawa-Sakata matrix, 
$U_\text{MNS}$~\cite{Maki:1962mu,Pontecorvo:1957cp},is nearly maximal.
$Z_3$ orbifolds offer several
possibilities for the generation of neutrino masses and mixings,
ranging from a purely Dirac formulation, to several implementations
of a type-I seesaw mechanism~\cite{seesaw:I}. Here we will
argue on the viability of each possibility.

This paper is organised as follows. In Section~\ref{yukawa} we conduct a brief
overview of the main properties of Yukawa couplings in this class of
$Z_3$ orbifold compactifications. In Section~\ref{higgsphenom}, 
we summarise the most relevant features of the extended Higgs sector.
The additional constraints on the orbifold parameters obtained from
reproducing the charged lepton masses are presented in 
Section~\ref{ch-lep-masses}, where we also discuss the tree-level
contributions to LFV processes. In
Section~\ref{orbifold_analysis}, we comment on the implications of the 
phenomenologically derived constraints regarding the 
properties of the compact space. Section~\ref{sec_neutrinos} is
devoted to the discussion of the viability of several mechanisms
regarding the generation of neutrino masses and mixings. We summarise
our findings in Section~\ref{concs}.

%
%
%
\section{Yukawa couplings in $\pmb{Z_3}$ orbifold models}\label{yukawa}

In this Section we briefly review the most relevant features of the
geometrical construction of the $Z_3$ orbifold
leading to the computation of the fermion mass matrices.

\begin{figure}[t]
  \begin{center} 
	\psfig{file=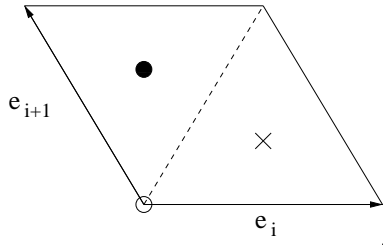,width=50mm,angle=0,clip=} 
    \caption{Two dimensional sublattice of the $Z_3$
    orbifold, symbolically denoting the fixed points as
    ($\circ,\bullet,\times$).}\label{fig:orbipicture}
\end{center}
\end{figure}

The construction is made by  dividing the 
$R^6$ space by a $[SU(3)]^3$
root lattice modded by the point group (P) with generator $\theta$,
where the action of $\theta$ on the lattice basis is 
$\theta e_i = e_{i+1}$, $\theta e_{i+1} = -(e_i + e_{i+1})$, with
$i=1,3,5$. The two-dimensional sublattices associated to $[SU(3)]^3$
are presented in Figure~\ref{fig:orbipicture}. 

In orbifold compactifications, twisted strings appear attached to fixed
points under the point group. In the case of the $Z_3$ orbifold there
are 27 fixed points under P, and therefore 27 twisted
sectors. We will denote the three fixed points of each two-dimensional
lattice as ($\circ,\bullet,\times$). 
The general form of the Yukawa couplings 
between the twisted fields in $Z_3$ orbifolds
is given by the Jacobi theta function
(see, for example, the Appendix of Ref.~\cite{Casas:1991ac}). 
These Yukawa couplings contain suppression factors that depend
on the relative positions of the fixed points to which the fields
involved in the coupling are attached, and on the size and shape of
the orbifold (i.e. the deformation parameters). Imposing invariance under 
the point group reduces these
parameters to nine: the three radii of the sublattices and the 
six angles between complex planes. The latter parameters correspond to the
VEVs of nine singlet fields appearing in the spectrum of the untwisted
sector, and which have perturbatively flat potentials. These so-called
moduli fields are usually denoted by $T$. In the most simple assumption, 
i.e. when the six angles are zero, there are only three relevant 
parameters which characterise the whole compact space.

Assuming that the two non-vanishing Wilson lines
correspond to the first and second sublattices, then the 27 twisted
sectors come in nine sets with three equivalent sectors in each
one. The three generations of matter (including Higgses) correspond to
changing the third sublattice component ($\circ,\bullet,\times$) of
the fixed point, while keeping the other two fixed. Consider for
example the following assignments of observable matter to fixed point
components in the first two sublattices,
\begin{align}
&Q \, \leftrightarrow \, \circ \,\circ\,
\quad \quad 
u^c \, \leftrightarrow \, \circ\,\circ\,
\quad \quad 
d^c \, \leftrightarrow \, \times\,\circ\,
\nonumber \\
&L \, \leftrightarrow \, \bullet\,\bullet\,
\quad \quad 
e^c \, \leftrightarrow \, \bullet\,\times\,
\quad \quad 
\nu^c \, \leftrightarrow \, \times\,\times\,
\nonumber \\
& \quad \quad \quad H^u \, \leftrightarrow \, \circ\,\circ\,
\quad \quad 
H^d \, \leftrightarrow \, \bullet\,\circ\,\quad \quad \quad \,.
\end{align}
In this case, 
the fermion
mass matrices before taking into account 
the effect of the FI
breaking, are given by the following expression~\cite{Abel:2002ih}:
\begin{eqnarray}\label{Z3mass:beforeFY}
\mathcal{M}^u= 
g \,N \,A^u\,, \quad \quad
\mathcal{M}^{d}= 
g \,N \,\varepsilon_1 \,A^d\,, \nonumber \\
\mathcal{M}^{\nu}= 
g \,N \,\varepsilon_1\,\varepsilon_3\,A^u\,, \quad \quad
\mathcal{M}^{e}= g \,N \varepsilon_3 \,A^d\,,
\end{eqnarray}
where
\begin{equation}\label{AuAd:bfFI}
A^{u}= \left(
\begin{array}{ccc}
w_2 & w_6 \,\varepsilon_5 & w_4 \,\varepsilon_5 \\
w_6 \,\varepsilon_5 & w_4 & w_2 \,\varepsilon_5 \\
w_4 \,\varepsilon_5 & w_2 \,\varepsilon_5 & w_6
\end{array}\right)\, , \quad \quad
A^{d}\left(
\begin{array}{ccc}
w_1 & w_5 \,\varepsilon_5 & w_3 \,\varepsilon_5 \\
w_5 \,\varepsilon_5 & w_3 & w_1 \,\varepsilon_5 \\
w_3 \,\varepsilon_5 & w_1 \,\varepsilon_5 & w_5
\end{array}\right)\,,
\end{equation}
$g$ is the gauge coupling constant, and $N$ is related to the 
volume of the $Z_3$ lattice unit cell such that $g\,N \approx 1$. 
In the above matrices 
$w_i$ denote the VEVs of the neutral components of the six Higgs 
doublet fields. Since we are assuming an orthogonal
lattice, i.e. with the six angles equal to zero, only the 
diagonal moduli $T_i$ (which are related to the radii of 
the three sublattices)
contribute to the Yukawa couplings, through the
suppression factors $\varepsilon_i$
\begin{equation}
\varepsilon_i \,\approx\,
3 \,e^{-\frac{2 \pi}{3} T_i}\,,\quad \quad i,j=1,3,5\,. 
\label{approximation}
\end{equation}

The existence of an anomalous $U(1)$ in the extended 
$SU(3) \times SU(2) \times U(1)^n$ gauge group generates a
Fayet-Iliopoulos $D$-term which could in principle break SUSY at
energies close to the string scale.
This term can be cancelled when 
scalar fields ($C_i$), which are singlets under $SU(3) \times SU(2)$,
develop large VEVs ($10^{16-17}$ GeV). The VEVs of these fields
($c_i$), have several important effects. Firstly, they 
break the original 
$SU(3) \times SU(2) \times U(1)^n$ gauge group down to the (MS)SM 
$SU(3) \times SU(2) \times U(1)$. Secondly, they induce very large
effective mass terms for many particles (vector-like triplets and doublets, as
well as singlets), which thus decouple from the low-energy
theory. Even so, the SM-like matter remains massless, surviving as the 
zero mass mode 
of combinations with the other (massive) states.
All these effects modify 
the mass matrices of the low-energy effective theory
(see Eq.~(\ref{Z3mass:beforeFY})), which, for the example studied
in~\cite{Abel:2002ih}, are now given by
\begin{align}\label{quark:mass}
\mathcal{M}^u  = \,g \,N \,a^{u^c} \,A^u \,B^{u^c}\,,& \quad \quad
\mathcal{M}^d  = \,g \,N \varepsilon_1 \,a^{d^c}\, A^d \,B^{d^c}\,, 
\nonumber \\
\mathcal{M}^{\nu} = \,g \,N \varepsilon_1\,\varepsilon_3 
\,a^{L}\,a^{\nu^c}\, B^{L}\,A^u \,B^{\nu^c}\,,& \quad \quad
\mathcal{M}^e  = \,g \,N \varepsilon_3 \,a^{L}\,a^{e^c}\,B^{L}\,A^d 
\,B^{e^c}\,,
\end{align}
where $A^{u,d}$
are the Higgs VEV matrices prior to FI breaking (see
Eq.~(\ref{AuAd:bfFI})), $a^{f}$ is given by 
\begin{equation}\label{af:def}
a^{f}\,=\,\frac{\hat c^f_2}{\sqrt{|\hat c_1^f|^2+|\hat c_2^f|^2}}\,,
\end{equation}
with $f=(u^c,d^c,L,e^c,\nu^c)$, and $B^f$ is the diagonal matrix defined as 
\begin{equation}\label{quark:B}
B^f\,=\, \operatorname{diag}\, (\,\beta^f \,\varepsilon_5,\, 1\,, 
\alpha^f/\varepsilon_5\,)\,. 
\end{equation}
Finally
\begin{align}\label{alpha:beta:def}
\alpha^f = \varepsilon_5 \sqrt{\frac{|\hat c_1^f|^2+|\hat c_2^f|^2}{
|\hat c_1^f \varepsilon_5|^2+|\hat c_2^f|^2}}\,, \quad \quad
\beta^f = \sqrt{\frac{|\hat c_1^f|^2+|\hat c_2^f|^2}{
|\hat c_1^f|^2+|\hat c_2^f \varepsilon_5|^2}}\,.
\end{align}
In the above, $\hat c_i^f$ are derived from the VEVs of the heavy fields
responsible for the FI breaking as 
\begin{equation}\label{hatc:def}
\hat c_1^f \,\equiv \,\varepsilon^{\prime (f)}\, c_1^f\,,\quad \quad 
\hat c_2^f \,\equiv \,\varepsilon^{\prime \prime (f)}\,c_2^f\,,
\end{equation}
where in each case $ \varepsilon^{\prime}$ and 
$ \varepsilon^{\prime \prime}$ can take any of the following values:
\begin{equation}\label{eprime:def}
\varepsilon^{\prime}\,, \,\, \varepsilon^{\prime \prime}
\equiv
1, \,\varepsilon_1, \,\varepsilon_3, \,\varepsilon_1 \,\varepsilon_3\,.
\end{equation}
Let us also stress that one should not take $\alpha^f$, $\beta^f$,
$\varepsilon_5$ and $a^f$ as independent parameters. 
In fact, Eqs.~(\ref{af:def},\ref{alpha:beta:def}) imply that 
\begin{equation}\label{af:alpha:beta}
a^f\,=\, 
\frac{\left( 1-{\alpha^f}^2\right)^{1/2}}{\alpha^f} \, 
\frac{\varepsilon_5}{\left( 1-\varepsilon_5^2 \right)^{1/2}}
\,=\,
\left( 1-\frac{1}{{\beta^f}^2}\right)^{1/2} \,
\frac{1}{\left( 1-\varepsilon_5^2 \right)^{1/2}}\,,
\end{equation}
so that for given values of $\varepsilon_5$ and $\alpha^f$, $\beta^f$
is fixed as
\begin{equation}\label{bf:alpha}
\beta^f\,=\,\frac{1}{\sqrt{1+\varepsilon_5^2
 \left(1-\frac{1}{{\alpha^f}^2} \right)}}\,.
\end{equation}
The most striking effect of the FI
breaking is that it enables the reconciliation of the Yukawa couplings
predicted by this scenario with experiment. 
In particular, as it has been shown \cite{EMT-2_06}, 
the quark spectrum and a successful Cabibbo-Kobayashi-Maskawa (CKM) matrix
can now be accommodated. 
Moreover, expanding the eigenvalues of the quark mass matrices up to leading
order in  $\varepsilon_5$, one can derive the following
relation
for the Higgs
VEVs in terms of the quark masses
\begin{align}\label{vev:quarkmass}
&\{w_1,w_3,w_5\} \,=
\frac{1}{g N \,\varepsilon_1 \,a^{d^c}}\,
\left\{
\frac{1}{\varepsilon_5 \beta^{d^c}} \left(m_d + \varepsilon_5^5
\frac{m_b^2}{m_s}\right), m_s, \frac{m_b \varepsilon_5}{\alpha^{d^c}}
\right\}\,, \nonumber\\
&\{w_2,w_4,w_6\} \, =
\frac{1}{g N \,a^{u^c}\,}
\left\{
\frac{1}{\varepsilon_5 \beta^{u^c}} \left(m_u + \varepsilon_5^5
\frac{m_t^2}{m_c}\right), m_c, \frac{m_t \varepsilon_5}{\alpha^{u^c}}
\right\}\,.
\end{align}
%
%
%
%
%
%
\section{The extended Higgs sector}\label{higgsphenom}
As it has been previously discussed, this
class of orbifold models contain naturally a replication of Higgs families. 
In this section we will summarise those features of the Higgs
sector relevant for the present analysis (for a more complete study
of generic SUSY models with three Higgs generations, 
see Ref.~\cite{EMT-1_06}).

By construction, let us consider an orbifold scenario containing three
generations of 
$SU(2)$ Higgs doublet superfields, with hypercharge $-1/2$ and $+1/2$, 
respectively coupling to down- and up- type quarks.
\begin{equation}\label{H:superf}
\widehat{H}_{1(3,5)}=
\left( \begin{array}{c}
\widehat{h}^0_{1(3,5)} \\
\widehat{h}^-_{1(3,5)}
\end{array} \right)\,,  \quad \quad
\widehat{H}_{2(4,6)}=
\left( \begin{array}{c}
\widehat{h}^+_{2(4,6)} \\
\widehat{h}^0_{2(4,6)}
\end{array} \right)\,.
\end{equation}
We assume the most general form of the superpotential for the quarks 
and leptons, which is given by 
\begin{align}\label{W:6Hdoublets}
W &=\,
\wh{Q}\, 
(Y_1^d\wh{H}_1+Y_3^d\wh{H}_3+Y_5^d\wh{H}_5)\wh{D}^c+
\wh{L}\,(Y_1^e\wh{H}_1+Y_3^e\wh{H}_3+Y_5^e\wh{H}_5)\wh{E}^c
 \nonumber \\ 
& +\, \wh{Q}\, (Y_2^u\wh{H}_2+Y_4^u\wh{H}_4+Y_6^u\wh{H}_6)\wh{U}^c + 
\wh{L}\, (Y_2^{\nu}\wh{H}_2+Y_4^{\nu}\wh{H}_4+Y_6^{\nu}\wh{H}_6)\wh{\nu}^c 
\nonumber \\
& +\mu_{12}\wh{H}_1\wh{H}_2+
\mu_{14}\wh{H}_1\wh{H}_4+\mu_{16}\wh{H}_1\wh{H}_6 
+ \,\mu_{32}\wh{H}_3\wh{H}_2+\mu_{34}\wh{H}_3\wh{H}_4 \nonumber \\ 
&
+\mu_{36}\wh{H}_3\wh{H}_6+\mu_{52}\wh{H}_5\wh{H}_2+\mu_{54}\wh{H}_5\wh{H}_4+
\mu_{56}\wh{H}_5\wh{H}_6\,,
\end{align}
where $\wh{Q}$ and $\wh{L}$ denote the quark and lepton $SU(2)_L$ 
doublet superfields, 
$\wh{U}^c$ and $\wh{D}^c$ are quark singlets, and
$\wh{E}^c$, $\wh{\nu}^c$, the lepton
singlet superfields. The Yukawa matrices $Y_i^f$
can be deduced from Eq.~(\ref{quark:mass}).
In what follows, we take
the $\mu_{ij}$ as effective parameters. 

The scalar potential receives the usual
contributions from $D$-, $F$- and SUSY soft-breaking terms, which we
write below, using for simplicity doublet components.
\begin{align}
V_F\,
=&
\operatornamewithlimits{\sum}_{\begin{smallmatrix}
{i,j=1,3,5}\\{l=2,4,6}
\end{smallmatrix}} \mu^*_{il}\, \mu_{jl}\, H_i^\dagger\, H_j
+
\operatornamewithlimits{\sum}_{\begin{smallmatrix}
{i=1,3,5}\\{k,l=2,4,6}
\end{smallmatrix}} \mu^*_{il} \,\mu_{ik}\, H_k^\dagger H_l \,,\nonumber\\
V_D\,
=&\,
\frac{g^2}{8} \,\operatornamewithlimits{\sum}_{a=1}^{3} 
\left[\, \operatornamewithlimits{\sum}_{i=1}^6
H_i^\dagger \,\tau^a\, H_i
\right]^2 + \,
\frac{g^{\prime 2}}{8} 
\left[\,\operatornamewithlimits{\sum}_{i=1}^6 \,
(-1)^i\, \left|H_i\right|^2 \,\right]^2\,,\nonumber\\
V_{\text{soft}}\,=&
\operatornamewithlimits{\sum}_{i,j=1,3,5}
(m^2_d)_{ij} \, H_i^\dagger\, H_j
+
\operatornamewithlimits{\sum}_{k,l=2,4,6}
(m^2_u)_{kl} \, H_k^\dagger H_l\,
-\operatornamewithlimits{\sum}_{\begin{smallmatrix}
{i=1,3,5}\\{j=2,4,6}
\end{smallmatrix}}
\left[(B\mu)_{ij}\, H_i\, H_j +\text{H.c.}\right]\,.\label{VDVFVS}
\end{align}
After electroweak (EW) symmetry breaking, the neutral components of the six
Higgs doublets develop VEVs, which we assume to be real, 
\begin{equation}
\vev{h^0_{1(3,5)}}\,=\,w_{1(3,5)}\,, 
\quad \quad \quad 
\vev{h^0_{2(4,6)}}\,=\,w_{2(4,6)}\,.
\end{equation}

For the purpose of minimising the Higgs potential and computing the
tree-level Higgs mass matrices, it proves more convenient to work in
the so-called ``Higgs basis''~\cite{Georgi:1978ri,Drees:1988fc}, 
where only two of the rotated fields develop VEVs:
\begin{align}\label{higgs:Ptransf}
\phi_i = P_{ij} h_j\,, \qquad \qquad\qquad \qquad\quad \quad\qquad \\
\vev{\phi^0_1} = \,  v_d\,=\,\sqrt{w_1^2+w_3^2+w_5^2}\,, \quad \quad
\vev{\phi^0_2} =\,  v_u\,=\,\sqrt{w_2^2+w_4^2+w_6^2}\,.
\end{align}
By construction, in order to comply with EW
symmetry breaking, the new VEVs must satisfy  
\begin{equation}\label{ewz2}
v_u^2+v_d^2 \,\approx \, (174\,\text{ GeV})^2\,,
\end{equation}
and we can now introduce a generalised definition for $\tan \beta$:
\begin{equation}\label{tb2}
\tan \beta = \frac{v_u}{v_d}\,.
\end{equation}
In the new basis, the free parameters at the EW scale are 
$m^2_{ij}$, $b_{ij}$, which has dimensions mass$^2$, 
and $\tan \beta$ (for a detailed discussion of the Higgs basis, including
the definition of the new parameters and of $P_{ij}$, 
we again refer the reader to~\cite{EMT-1_06}), and the minimisation
equations simply read:
\begin{equation}\label{minima:du}
\begin{array}{ll}
m^2_{11}\,=\,\,
b_{12} \,\tan \beta - \frac{M_Z^2}{2}\,\cos 2 \beta\,, \quad \quad
\quad \quad \quad
&
m^2_{22}\,=\,\,
b_{12} \,\cot \beta + \frac{M_Z^2}{2}\,\cos 2 \beta\,, \\
m^2_{13}\,=\,\,
b_{32} \,\tan \beta \,,
&
m^2_{24}\,=\,\,
b_{14}\, \cot \beta\,, \\
m^2_{15}\,=\,\,
b_{52} \, \tan \beta \,,
&
m^2_{26}\,=\,\,
b_{16} \,\cot \beta \,.
\end{array}
\end{equation}
The Higgs spectrum derived from the tree-level 
potential and the previous conditions
is far richer than in the usual MSSM case. The neutral sector consists
of six CP-even and five CP-odd scalars, while the charged sector
will contain ten mass eigenstates. This provides not only a more challenging
scenario for potential Higgs detection, but also allows the appearance of 
dangerous tree-level FCNCs which are severly constrained by observation. 
The phenomenology related to this multiple-Higgs model and the effects on
quark flavour changing transitions have been previously studied 
in~\cite{EMT-2_06}. The impact on lepton flavour violating processes
will be discussed in Section~\ref{LFV:section}. 

%
%
%
\section{Charged leptons}\label{ch-lep-masses}
%
%
As mentioned in the Introduction, the previous analysis~\cite{EMT-2_06}
of the orbifold parameter space has already
severely constrained the free parameters of the orbifold. 
As discussed, we have verified that one could successfully
reproduce the observed hierarchy and mixings in the quark sector, and
avoid potentially dangerous tree-level FCNCs with a fairly light Higgs
boson spectrum. 
In what follows, we extend our analysis to the lepton sector. In this
section we address how reproducing the charged lepton masses further 
constrains the orbifold parameters, and also discuss possible
tree-level lepton flavour violation, arising from the exchange of
neutral Higgses.

\subsection{Charged lepton masses}\label{sec_chlepmasses}
We start by considering the mass matrix for the charged 
leptons, which after FI breaking is given by\footnote{Note 
that the expression for the matrix ${\cal M}^{e}$
in Eq.~(\ref{ch-leplon_mass_matrix}) corrects the misprint in
Ref.~\cite{Abel:2002ih}, Eq. (69), where the matrix product was taken 
in the order $ABB$. A similar correction for neutrino masses will be
subsequently taken
into account in Eqs.~(\ref{finalnudirac}) and (\ref{finalnuda}).} 
\bea
{\cal M}^{e}\,=\,g\,N\,\epsilon_3\,a^L\,a^{{e}^c}\, B^L\, A^{d}\, B^{e^c} 
\,=\,g\,N\,\epsilon_3\,a^L\,a^{{e}^c}\,
\left( \begin{array}{ccc}
v_1 \,\varepsilon_5^2 \,\beta^{L}\,\beta^{e^c} & 
v_5\,\varepsilon_5^2\,\beta^L  & v_3\,\epsilon_5\,\alpha^{e^c}\,\beta^L \\
v_5\,\varepsilon_5^2\,\beta^{e^c}  & v_3 & 
v_1\,\alpha^{e^c} \\
v_3\,\varepsilon_5\,\alpha^{L}\, \beta^{e^c}  & v_1\,\alpha^L  & 
v_5\,\alpha^{L}\,\alpha^{e^c}/\varepsilon_5^2
\end{array}\right)\,,
\label{ch-leplon_mass_matrix}
\eea
where $A^d$, $a^{L,e^c}$ and $B^{L,e^c}$ have been defined in 
Eqs.~(\ref{AuAd:bfFI}, \ref{af:def}-\ref{alpha:beta:def}), 
setting $f=L, e^c$.
The next step in the analysis is to determine whether one can find
regions of the parameter space where the charged lepton masses can be
obtained. We recall that most of the variables appearing
in Eq.~(\ref{ch-leplon_mass_matrix}), namely $g$, $N$, $\varepsilon_5$ 
and the Higgs VEVs $w_i$ are also related to the quark sector
of the model, and are thus already tightly constrained~\cite{EMT-2_06}.
We consider the quark input sets studied in Ref.~\cite{EMT-2_06},
used to fix the six Higgs VEVs, and which
correctly reproduce the correct mass spectrum for both the up- and
down-quarks $\{m_u,\,m_d,\,m_c,\,m_s,\,m_t,\,m_b\}$:  
\bea
\textrm{SET
  A}=\{0.0040,\,0.008,\,1.35,\,
0.130,\,180,\,4.40\}  
\ \textrm{GeV},
\label{setAquark} \\
\textrm{SET
  B}=\{0.0035,\,0.008,\,1.25,\,
0.100,\,178,\,4.50\}  
\ \textrm{GeV},
\label{setBquark} \\
\textrm{SET
  C}=\{0.0035,\,0.004,\,1.15,\,
0.080,\,176,\,4.10\}  
\ \textrm{GeV},
\label{setCquark} \\
\textrm{SET
  D}=\{0.0040,\,0.006,\,1.20,\,
0.105,\,178,\,4.25\}  
\ \textrm{GeV},
\label{setDquark}
\eea
and scan over the $\varepsilon_5$, $\alpha^{u^c}$ and $\alpha^{d^c}$
intervals compatible with realistic quark masses and mixings,
\be
0.0085 \,\leq \,\varepsilon_5\, \leq \,0.0260\,,\qquad \,0.040 \,\leq
\,\alpha^{u^c}\, 
\,\leq 0.370\,,\qquad \,0.190\, \leq\, \alpha^{d^c} \,\leq \,0.842\,, 
\label{setBquark_eps}
\ee

The value of $\tan{\beta}$, which is also crucial, is tightly related
to $\varepsilon_1$~\cite{EMT-2_06}. 
From the previous values, and employing Eqs.~(\ref{vev:quarkmass}), 
(\ref{ewz2}) and (\ref{tb2}), we can also derive the value of $gN$,
which is obtained from the following expression:
\begin{align}\label{gNorbifold}
g\,N &= \frac{1}{a^{u^c}}\,
\frac{\left( 1+\tan^2 \beta \right)^{1/2}}{\tan \beta}\,\,
\frac{\sqrt{\frac{1}{(\varepsilon_5 \,\beta^{u^c})^2} \left(m_u +
    \varepsilon_5^5 
\frac{m_t^2}{m_c}\right)^2+
m_c^2 +  \left(\frac{m_t \,\varepsilon_5}{\alpha^{u^c}}\right)^2
}}{174\, \text{GeV}}\,.
\end{align}
Thus, once the several orbifold parameters are determined, one can
derive information on the intrinsic orbifold properties, such as the value
of the orbifold normalisation constant $N$, or the heterotic coupling
constant $g$. The numerical analysis of this subsection is
instrumental in obtaining the latter information.

\begin{figure}
  \begin{center} \hspace*{-10mm}
    \begin{tabular}{cc}
	\psfig{file=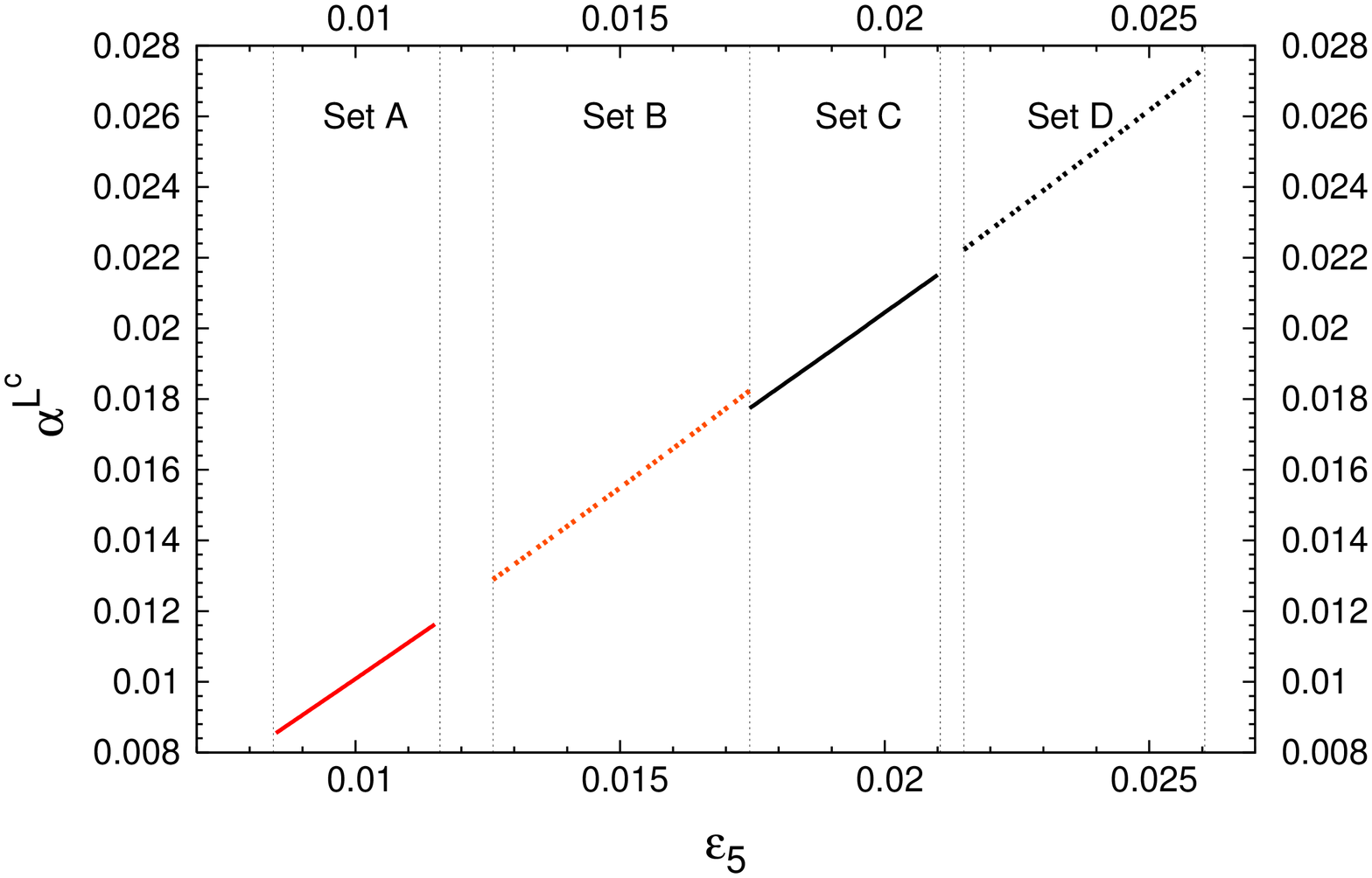,width=55mm,angle=270,clip=} &
	\psfig{file=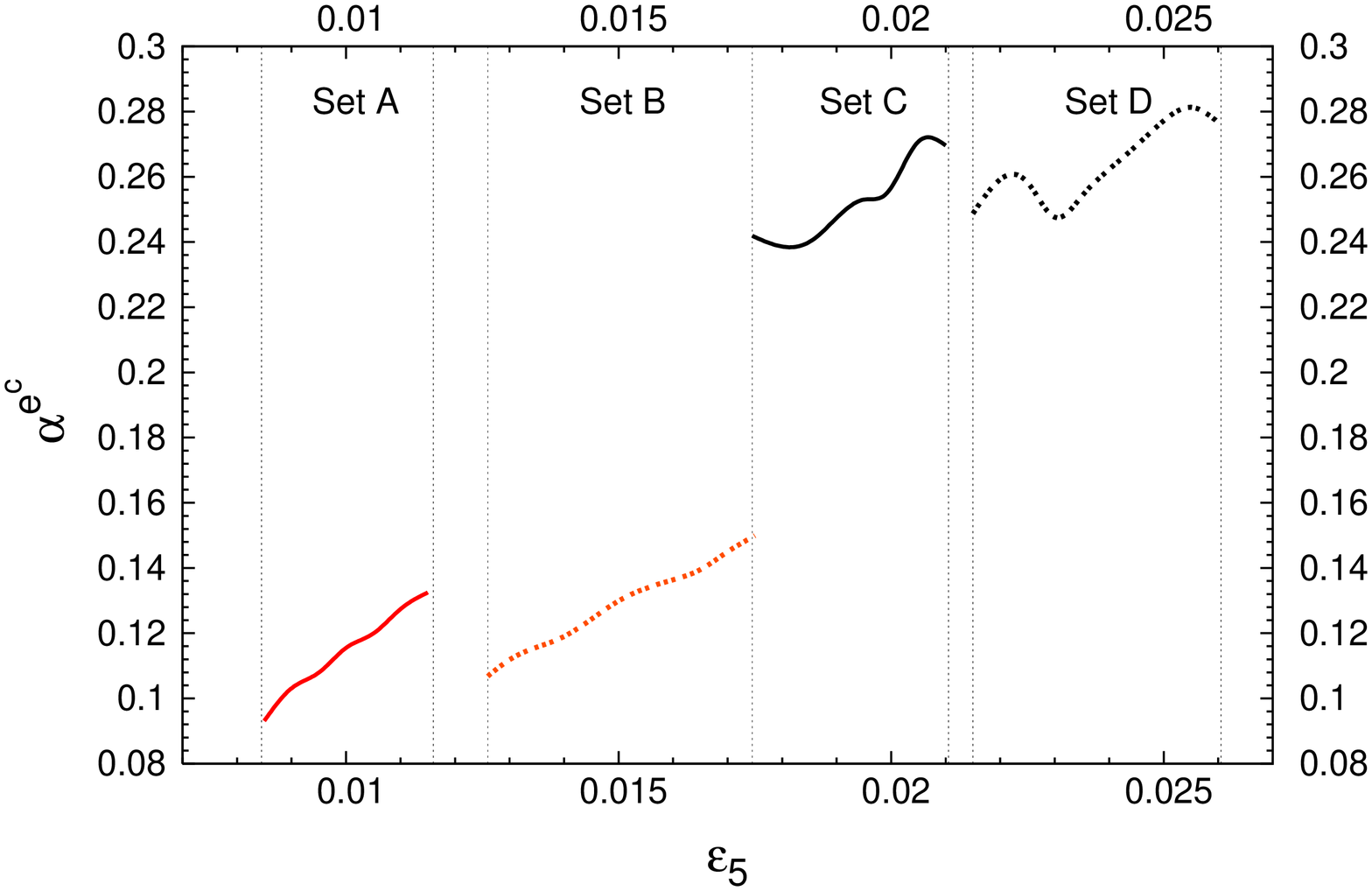,width=55mm,angle=270,clip=} 
    \end{tabular}
    \caption{Correlation between the orbifold parameters
($\alpha^{L^c}$, $\varepsilon_5$) and ($\alpha^{e^c}$,
    $\varepsilon_5$), for $\tan \beta=5$ and the distinct sets of 
      input quark masses, A-D
      (red full lines, red dashed lines, full black and dashed black
      lines, respectively).} 
    \label{fig:orbifold:alpha:Le:e5}
  \end{center}
\end{figure}

\begin{figure}
  \begin{center} 
\psfig{file=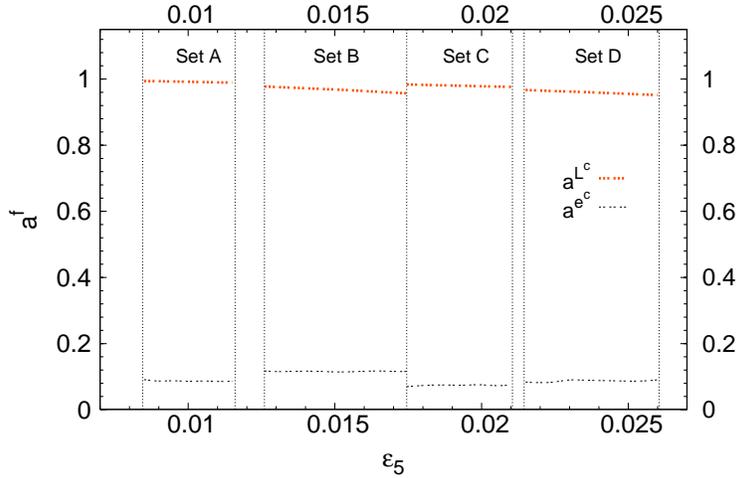,width=70mm,angle=270,clip=} 
    \caption{Correlation between the orbifold parameters
$a^{L^c}$, $a^{e^c}$ and
    $\varepsilon_5$, for the quark sets A-D and $\tan{\beta}=5$.} 
    \label{fig:orbifold:aLae:e5}
  \end{center}
\end{figure}

Having set the quark parameters and choosing $\tan{\beta}=5$ as an
example, we proceed to determine 
$\varepsilon_3$, $\alpha^L$, and $\alpha^{e^c}$. These values will in
turn allow to derive the mass matrix for the charged leptons. In
agreement with experimental data~\cite{pdg2004}, the latter 
eigenvalues should be
\bea
\{m_e,\,m_\mu,\,m_\tau\}=\{0.511,\,105.41,\,1778.45\}\ \textrm{MeV}.
\label{chlept_masses}
\eea
In Figures~\ref{fig:orbifold:alpha:Le:e5} and \ref{fig:orbifold:aLae:e5} 
we present the values of $\alpha^{L,e^c}$ and $a^{L,e^c}$ giving
rise to the correct masses for both the quark and the charged-lepton sectors. 
The scan over $\varepsilon_5$ has been conducted for the four quark sets in 
Eqs.~(\ref{setAquark}\,-\ref{setDquark}). 
We can see that the behaviour of $\alpha^{L,e^c}$ is completely
analogous to what had been observed for the quark
sector~\cite{EMT-2_06}, which is not unexpected, given that the Yukawa
couplings for the charged leptons closely follow those of the
down-type quarks.

Let us comment on the suppression factor $\varepsilon_3$. 
As seen from Eq.~(\ref{ch-leplon_mass_matrix}), $\varepsilon_3$ is
a global factor in the charged-lepton mass matrix. 
This allows its value to be modified without affecting the mass
eigenstates, provided that $\tan \beta$ (i.e. the ratio of the Higgs
VEVs) is accordingly changed.
In other words, $\tan \beta$ is still an unconstrained degree of
freedom, a fact that is particularly useful for the analysis involving the
Higgs sector (as discussed in~\cite{EMT-1_06}).
\begin{figure}
  \begin{center} 
\psfig{file=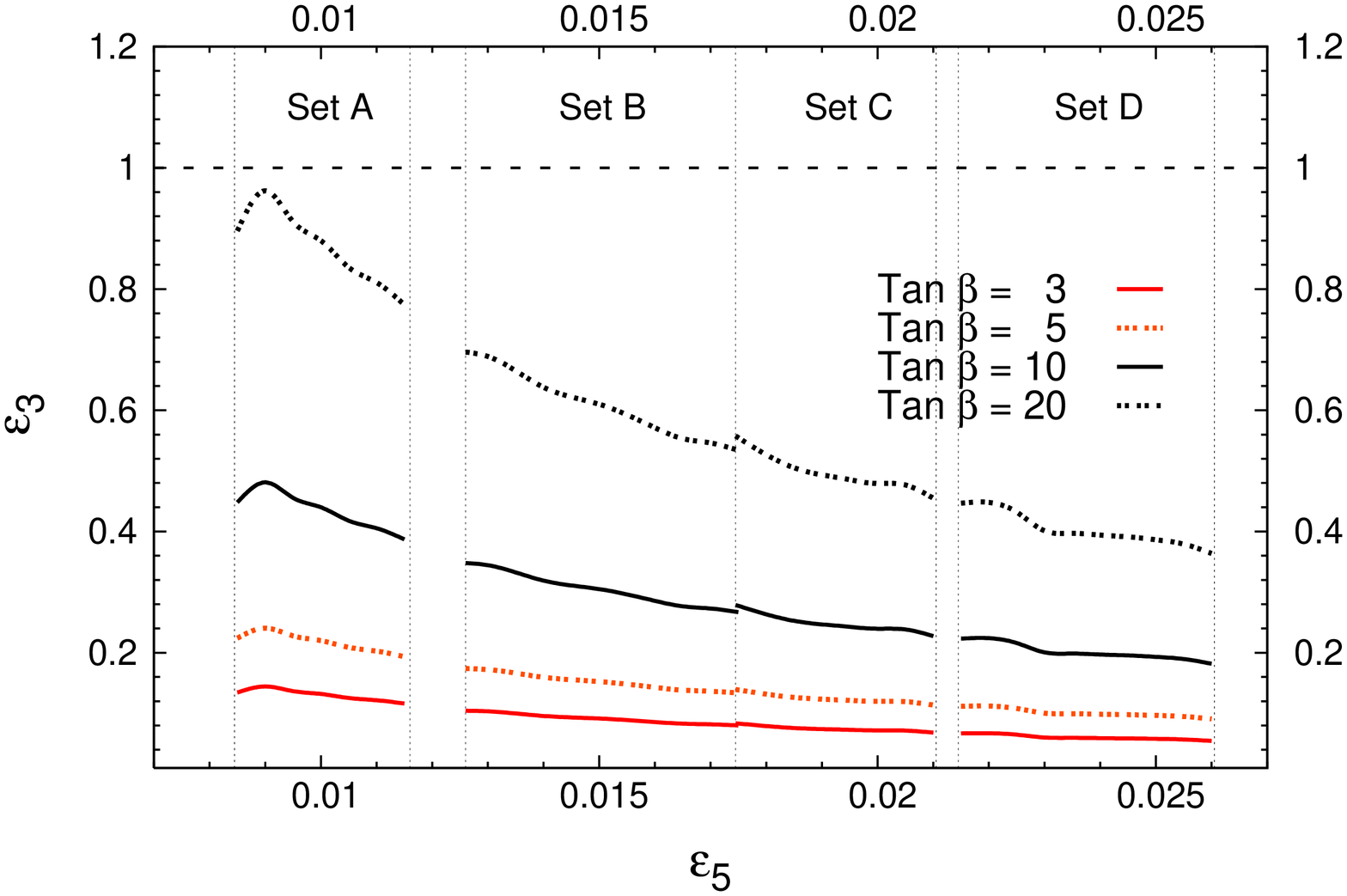,width=70mm,angle=270,clip=} 
    \caption{Correlation between $\varepsilon_3$ and $\varepsilon_5$
    for different values of $\tan \beta$.} 
    \label{fig:orbifold:e3:e5}
  \end{center}
\end{figure}
In Figure~\ref{fig:orbifold:e3:e5} we display the relation between 
$\varepsilon_3$ and $\varepsilon_5$ for four different values of $\tan \beta$. 
It is worth mentioning here that, given a particular quark input set,
the value of $\tan \beta$ is bounded from above
in order to avoid $\varepsilon_3 > 1$.
For example, for set A this bound is close to $\tan \beta=20$. In
fact, and as already noticed in the study of the quark sector, the
phenomenological viability of these orbifold constructions favours
lower values of $\tan \beta$ (not only based on reproducing a viable
spectrum, but also related with avoiding excessive FCNCs).

Not only can the quantities $\varepsilon_i$ be understood as suppression
factors which affect the Yukawa couplings
(providing the desired mass hierarchy between fermions),
but they are also subject to perturbativity constraints. Concerning
the latter, for example $\varepsilon_3$ is actually given by~\cite{Abel:2002ih}
\bea
\varepsilon_3\,=\, 3\, e^{-\frac{2\,\pi}{3}\,T_3}
(1+6\,e^{-2\,\pi\, T_1}+6\,e^{-2\,\pi\, T_5}+...)\,
\approx \,3\, e^{-\frac{2\,\pi}{3}\,T_3}\,,
\label{epsilon1}
\eea
where the last approximation corresponds to the assumption of
Eq.~(\ref{approximation}). Clearly, if $\varepsilon_i$ are in general large, 
$T_i$ have to be small, and therefore perturbativity is spoiled.
Under the approximation of Eq.~(\ref{approximation}), one can write
\bea
T_i\, =\, -\frac{3}{2\,\pi}\,\ln{\frac{\varepsilon_i}{3}}\,,
\label{Tfromeps}
\eea
and, as a consequence, we verify that $\varepsilon_i$ cannot be larger than 3, 
since the $T_i$ VEVs are proportional to $R_i^2$, and therefore 
positive. From the analysis of the orbifold parameters, 
one can obtain useful information about the high-energy 
configuration of the string model, namely the size and properties of 
the compact space,
as well as its relation with the gauge unification scale.
The allowed regimes for the three $T_i$ and their physical implications 
will be studied in detail in Section~\ref{orbifold_analysis}.
%
%
%
%
\subsection{Tree-level lepton flavour violation}\label{LFV:section}

Identical to what occurs for the quark sector, having a model with
Higgs family replication opens the possibility of
tree-level FCNCs in the lepton sector, 
contrary to what occurs in the SM or in the MSSM.
Given the fact that flavour-violating interactions are very suppressed
in Nature, one should ensure that the present model does not induce
excessively large contributions to these processes.
In a general multi-Higgs model, it is widely recognised that the most
stringent bounds arise from the smallness of the masses of the 
long- and short-lived neutral kaons. It has been previously 
verified~\cite{EMT-2_06} that for a relatively light Higgs boson 
spectrum of order $\thicksim 1-5$ TeV, the present orbifold model 
is in very good agreement with experimental data. 
The analysis was also extended to the $B$- and $D$-meson systems,
leading to similar bounds for the Higgs masses.
With the inclusion of the charged lepton sector in our analysis, it is
only natural to expect dangerous lepton flavour-violating
interactions.
Regarding these interactions, here we have focused on the 
branching ratios (BRs) of pure leptonic decays of the type 
$l_i\to 3l_j$, which have been identified in the literature as the
less suppressed processes~\cite{MWL-80,Shanker-81}. In the context of
the present orbifold model, these decays are going to be generated 
by Yukawa interactions mediated by neutral Higgs
bosons\footnote{We stress here that there are no tree-level
  contributions to other LFV processes, like
  radiative decays of the type $l_i\to l_j \gamma$, which only occur
  at one-loop level.}. As shown in recent studies of LFV in SUSY
models with one Higgs
family~\cite{Arganda:2005ji,Antusch:2006vw,Arganda:2007jw}, 
the one-loop contributions to flavour
violating processes can be extremely large for sizable values
of $\tan \beta$ and a Higgs mass of order 100\,-150 GeV.
In our case, and as will be shortly confirmed, the requirement that
the Higgs bosons are heavy enough to suppress the dangerous quark FCNC
interactions indeed ensures that the leptonic processes remain 
several orders of magnitude below the respective experimental bounds.

\begin{figure}
  \begin{center} 
\psfig{file=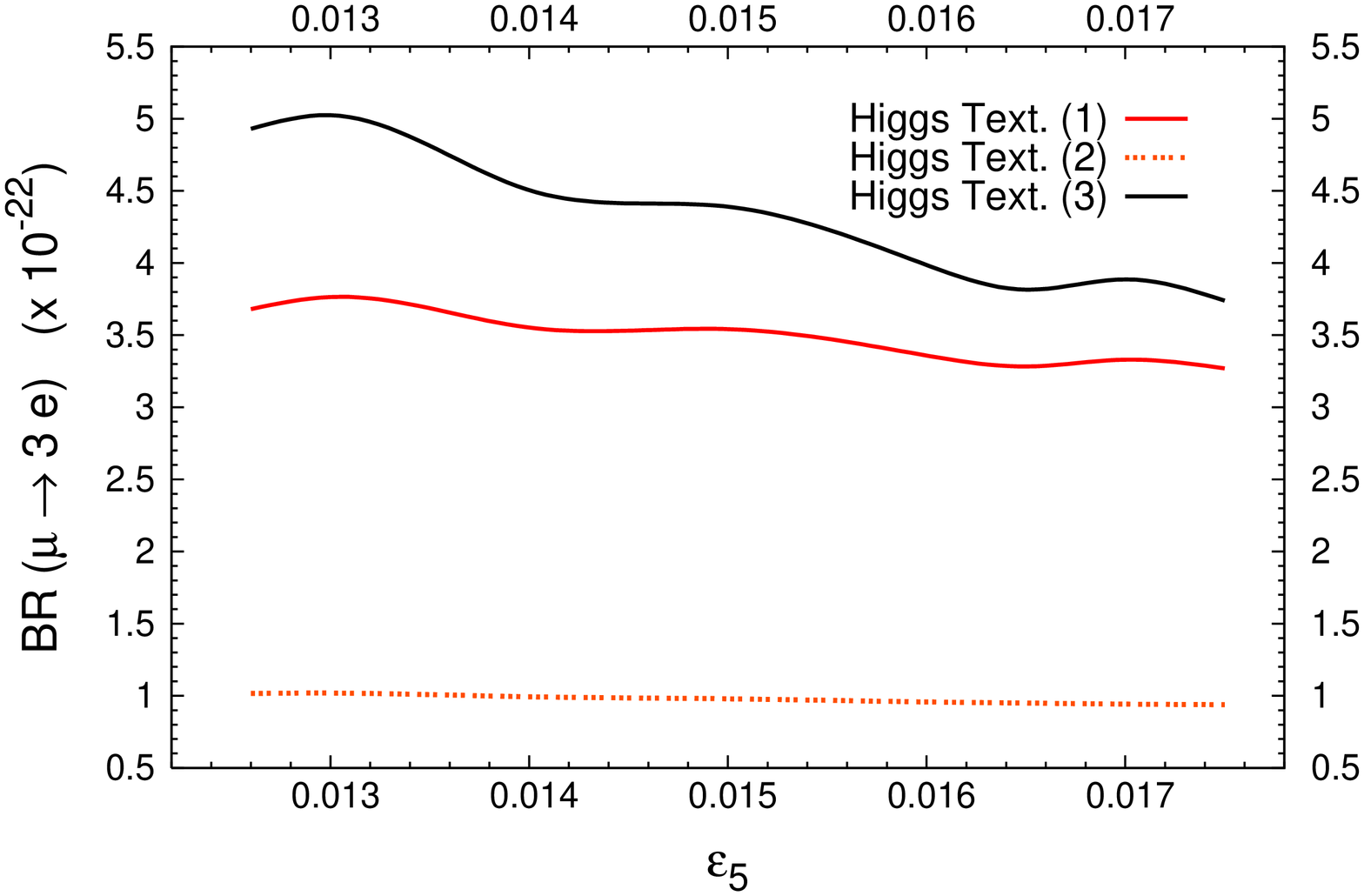,width=70mm,angle=270,clip=} 
    \caption{BRs for the Higgs-mediated $\mu \to 3 e$ decay as a
    function of $\varepsilon_5$, for quark set A and Textures (1-3).} 
    \label{fig:orbifold:BR:mu:eee:e5}
  \end{center}
\end{figure}

\begin{figure}
  \begin{center} 
\psfig{file=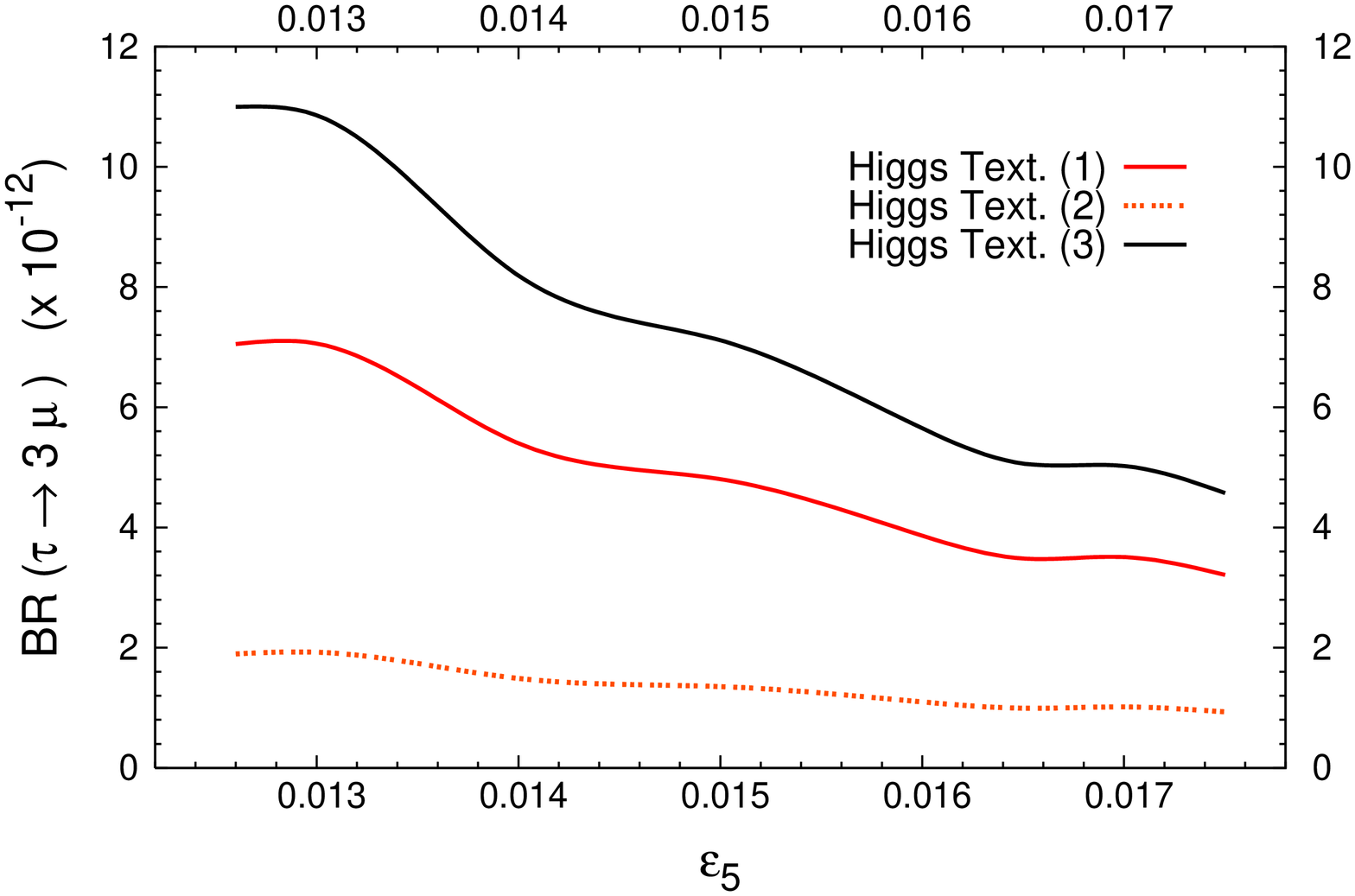,width=70mm,angle=270,clip=} 
    \caption{BRs for the Higgs-mediated $\tau \to 3 \mu$ decay as a
    function of $\varepsilon_5$, for quark set A and Textures (1-3).} 
    \label{fig:orbifold:BR:tau:mumumu:e5}
  \end{center}
\end{figure}

\begin{figure}
  \begin{center} 
\psfig{file=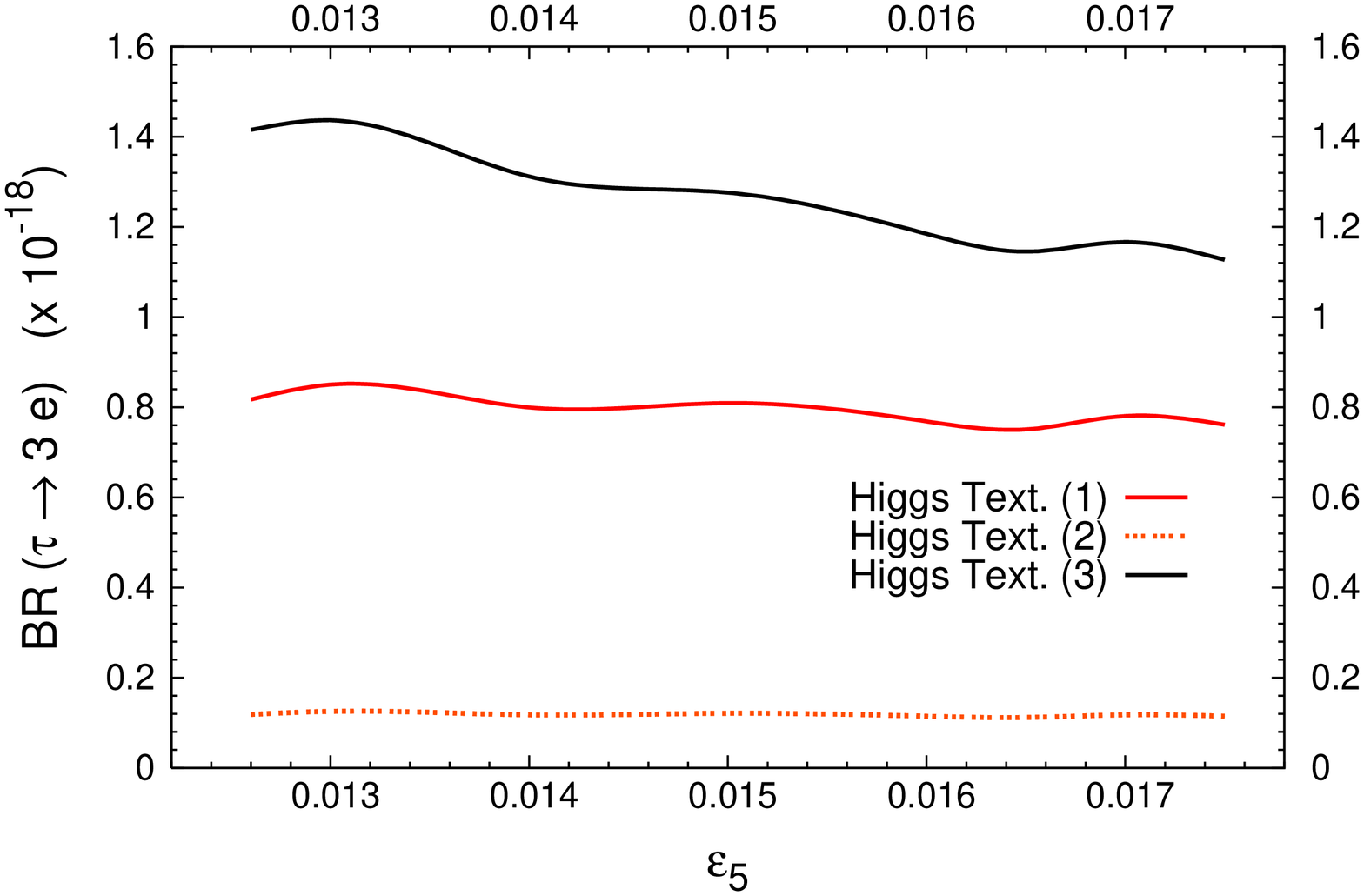,width=70mm,angle=270,clip=} 
    \caption{BRs for the Higgs-mediated $\tau \to 3 e$ decay as a
    function of $\varepsilon_5$, for quark set A and Textures (1-3).} 
    \label{fig:orbifold:BR:tau:eee:e5}
  \end{center}
\end{figure}

In order to study the occurrence of tree-level LFV in the
charged-lepton sector we consider the branching ratios of
three-body decays, $l_i\to 3l_j$, mediated by a neutral physical Higgs
eigenstate ($\varphi_k$). 
The transition amplitudes and BRs of these processes are then given by
%
\be
\Gamma{(l_i\,\to\,  3l_j|\varphi_k)}\,=\,
\frac{|\mathcal{Y}^k_{ji}|^2\,|\mathcal{Y}^k_{jj}|^2}{128\,
m_{\varphi_k}^4}\,
\frac{m_{l_i}^5}{192\,\pi^3}\,
\quad \quad
\mathrm{BR}{(l_i\to 3l_j |\varphi_k)}\,
=\,\frac{1}{128\,G_F^2}\,\frac{|\mathcal{Y}^k_{ji}|^2\,
|\mathcal{Y}^k_{jj}|^2}
{m_{\varphi_k}^4}
\ ,
\label{lepton-branching ratios}
\ee
%
where $m_{l}$ is the lepton mass, $m_{\varphi_k}$ the mass of the mediating
scalar/pseudoscalar neutral Higgs, and $\mathcal{Y}^k_{ij}$ is the $i,j$
element 
of the Yukawa coupling matrix, in the physical mass-eigenstate basis, 
defined as
\be\label{calY}
\mathcal{Y}_{ij}^{k} =
(S^\dagger)_{kl} \, \,
(V_R^e\, Y^{e}_l \,V^{e\dagger}_L)_{ij}\,.
\ee
In the above, the Higgs physical states
are related to the original interaction eigenstates by
$\varphi_k=S_{kl}\,h^0_l$,
where $h^0_l$ are the neutral components of the Higgs doublets
(see Eq.~(\ref{H:superf})). $V_R^e$ and $V_L^{e\dagger}$ are the matrices
which diagonalise the charged-lepton mass matrix, and $Y^e_l$ 
(with $l=1,2,3$) are the 
three charged-lepton Yukawa matrices associated to the down-type Higgses,
as shown in Eq.~(\ref{W:6Hdoublets}).
Using the above expressions, we can now compute the contributions of
the full Higgs spectrum (six scalars and five pseudoscalars) to the
LFV decays. To do so, we choose three distinct Higgs mass textures,
already considered in a previous study~\cite{EMT-2_06}.
Working in the Higgs basis (see Section~\ref{higgsphenom})
these can be summarily defined via the following parametrisation,
which allows to define the Higgs sector
via six dimensionless parameters as 
\begin{equation}\label{higgs:tx:mud}
m^{(d)}_{ij}\, =\, \left(
\begin{array}{ccc}
\otimes&\otimes  &\otimes  \\
\otimes & x_3 & y \\
\otimes & y & x_{5}
\end{array}\right)\times {1 \text{TeV}}\,,
\quad 
m^{(u)}_{kl}\,=\, \left(
\begin{array}{ccc}
\otimes&\otimes  &\otimes  \\
\otimes & x_4 & y \\
\otimes & y & x_{6}
\end{array}\right)\times {1 \text{TeV}}\,,
\quad
\sqrt{b_{ij}}\,=\,b \times {1 \text{TeV}}\,.
\end{equation}
In the above, $m^{d(u)}_{ij}$ should be understood as the $i,j=1,3,5$
($k,l=2,4,6$) submatrices of the $6\times6$ matrix that encodes the rotated
soft-breaking Higgs masses in the Higgs basis (see~\cite{EMT-1_06}).
The symbol $\otimes$ denotes an entry which is
fixed by the minima conditions of Eq.~(\ref{minima:du}). 
For the Yukawa matrices, we will employ
the quark Set A of Eq.~(\ref{setAquark}) and those values of 
$\varepsilon_5$, $\alpha^{L^c}$ and $\alpha^{e^c}$ compatible
 with realistic masses for the charged leptons
(as analysed in Section~\ref{ch-lep-masses}). 
Other sets for the quark masses, B, C or D, will lead to similar results.
For simplicity, we will take a near-universality limit for the Higgs-sector
textures introduced in Eq.~(\ref{higgs:tx:mud}). 
Regarding the value of $\tan \beta$,
and unless otherwise stated, we shall take $\tan \beta=5$ in the subsequent
analysis. We consider the following three cases, with the associated tree-level
scalar and pseudoscalar Higgs spectra:
\begin{itemize}
\item[(1)]
$x_3=x_4=0.5\,, x_5=x_6=0.75\,, y=0.1\,, b=0.1\,$

$m^s = \{82.5, 190.6, 493.9, 515.9, 744.4, 760.2\}$ GeV\,;

$m^p = \{186.8, 493.9, 515.9, 744.4, 760.2\}$ GeV\,.

\item[(2)] 
$x_3=x_4=0.75\,, x_5=x_6=1, y=0.25\,, b=0.2\,$

$m^s = \{83.6, 292.9, 733.6, 785.9, 987.6, 1057.0\}$ GeV\,;

$m^p = \{291.1, 733.6, 785.9, 987.6, 1057.0\}$ GeV\,.
\item[(3)]
$x_3=x_4=0.5\,, x_5=5\,, x_6=7.5, y=0.5\,, b=0.1\,$

$m^s = \{82.7, 201.4, 492.4, 516.4, 5000, 7500\}$ GeV\,;

$m^p = \{197.9, 492.4, 516.4, 5000, 7500\}$ GeV\,.
\end{itemize}
In the above, $m^s$ and $m^p$ respectively denote 
the values for the physical scalar and pseudoscalar masses.
The results for the decays $\mu \to 3 e$, $\tau \to 3 \mu$ and $ \tau
\to 3 e$ are summarised in Figs.~\ref{fig:orbifold:BR:mu:eee:e5}, 
\ref{fig:orbifold:BR:tau:mumumu:e5} and~\ref{fig:orbifold:BR:tau:eee:e5}.
From the latter, we immediately observe that the Higgs-mediated
contributions to the $l_i\to 3l_j$ branching ratios always lie several
orders of magnitude below the experimental limits (collected in
Table~\ref{LFV:bounds:future}).
This occurs even for Texture (1),
associated with a spectrum containing only light 
(below 1 TeV) Higgs particles.
\begin{center}
\begin{table}\hspace*{45mm}
\begin{tabular}{|c|c  c |}
\hline
LFV process & Present bound & Future sensitivity \\
\hline
BR($\mu \to 3\,e$) & $1.0 \times 10^{-12}$  & 
$10^{-13}$  \\
BR($\tau \to 3\,e$) & $2.0 \times 10^{-7}$  & 
$10^{-8}$  \\
BR($\tau \to 3\,\mu$) & $1.9 \times 10^{-7}$  & 
$10^{-8}$  \\\hline
\end{tabular}
\caption{Present bounds and future sensitivities for the LFV 
processes \cite{Bellgardt:1987du,Aubert:2003pc,Akeroyd:2004mj}.}
\label{LFV:bounds:future}
\end{table}
\end{center}
Regarding other relevant LFV processes, as for example 
leptonic conversion processes in heavy nuclei, which could in principle also
receive important tree-level contributions, we have not discussed them
here, as these conversion processes are always assumed to be of the same 
order or even sub-dominant  with respect to the leptonic decays (see, 
for example,~\cite{MWL-80}, \cite{Shanker-81} or \cite{Ng:1993ey}).
The extremely low contribution to the purely leptonic decays previously studied
(between 5-10 orders of magnitude below the present experimental bounds) 
renders the impact of these LFV processes clearly negligible, when  
compared to the flavour-changing processes occurring in the quark
sector.

%
%
%
\section{Orbifold analysis at the string scale}\label{orbifold_analysis}
With the full determination of the quark and charged-lepton sectors we
are now ready to address the implications of imposing phenomenological
viability at the string level. 
As previously discussed, the characterisation of the orbifold model is
tightly related to the determination of the geometrical suppression
factors, $\varepsilon_i$, which in turn are instrumental in complying
with the different fermion mass hierarchies. As we will discuss in 
Section~\ref{sec_neutrinos}, $\varepsilon_i$ will further affect the
neutrino Yukawa couplings, with an important impact on the seesaw scale.
At this point, it is also relevant to mention that 
we will not take into account the effect of the
renormalisation group equations (RGE) on the quark 
and lepton mass matrices presented 
in the previous sections. The flavour structure for the masses
is associated with a mechanism taking place at a very
high energy scale. However, and given the clearly hierarchical
structure of the mass matrices, one does not expect that 
RGE running will significantly affect the predictions of the model.

In this section, we briefly comment on the information  
about the shape and size of the compact space, and also discuss the hints 
on the gauge properties of the string model, which can be 
inferred from the already constrained values of $\varepsilon_i$.

Let us firstly consider the value of the product of the heterotic
coupling constant, $g$, by the orbifold normalisation constant $N$, 
defined in Eq.~(\ref{gNorbifold}). The normalisation constant is given
by 
\bea
N\,=\,{\sqrt V}\, \frac{3^{3/4}}{8\,\pi^3}\,
\frac{\Gamma^6 (\frac{2}{3})}{\Gamma^3 (\frac{1}{3})}\,.
\label{normalization}
\eea
In the latter, $V$ denotes the volume of the unit cell of the $Z_3$ lattice,
\bea
V\,=\,(R_1\,R_3\,R_5)^2\,\left(\sin{\frac{2\,\pi}{3}}\right)^3\,
\label{unitcellvolume}
\eea
where $R_{1,3,5}$ are the unit cell radii in each sublattice, defined
in terms of the three $T$-moduli as 
\bea
R_i\,=\,\frac{4\,\pi}{3^{1/4}}\,\sqrt{\,\textrm{Re}\,T_i}\,.
\label{RfromT}
\eea
From Eq.~(\ref{gNorbifold}), taking only the dominant
terms into account, we can verify that the assumption of 
$g\,N\,\approx 1$~\cite{Abel:2002ih} is indeed valid for values
of $\varepsilon_5$, $\alpha^{u^c}\ll 1$ and $\tan \beta \geq 3$, since
\be
g\,N\, \approx\, \frac{\varepsilon_5}{\alpha^{u^c}\,a^{u^c}}\,
\frac{(1+\tan^2 \beta)^{1/2}}{\tan \beta}\,
\frac{m_t}{174\, \text{GeV}}
\,\approx\, 
\frac{(1-\varepsilon_5)^{1/2}}{(1-\alpha^{u^c})^{1/2}}\,\approx\, 1\,,
\label{gNlimit}
\ee
\begin{figure}
  \begin{center} \hspace*{-10mm}
    \begin{tabular}{cc}
	\psfig{file=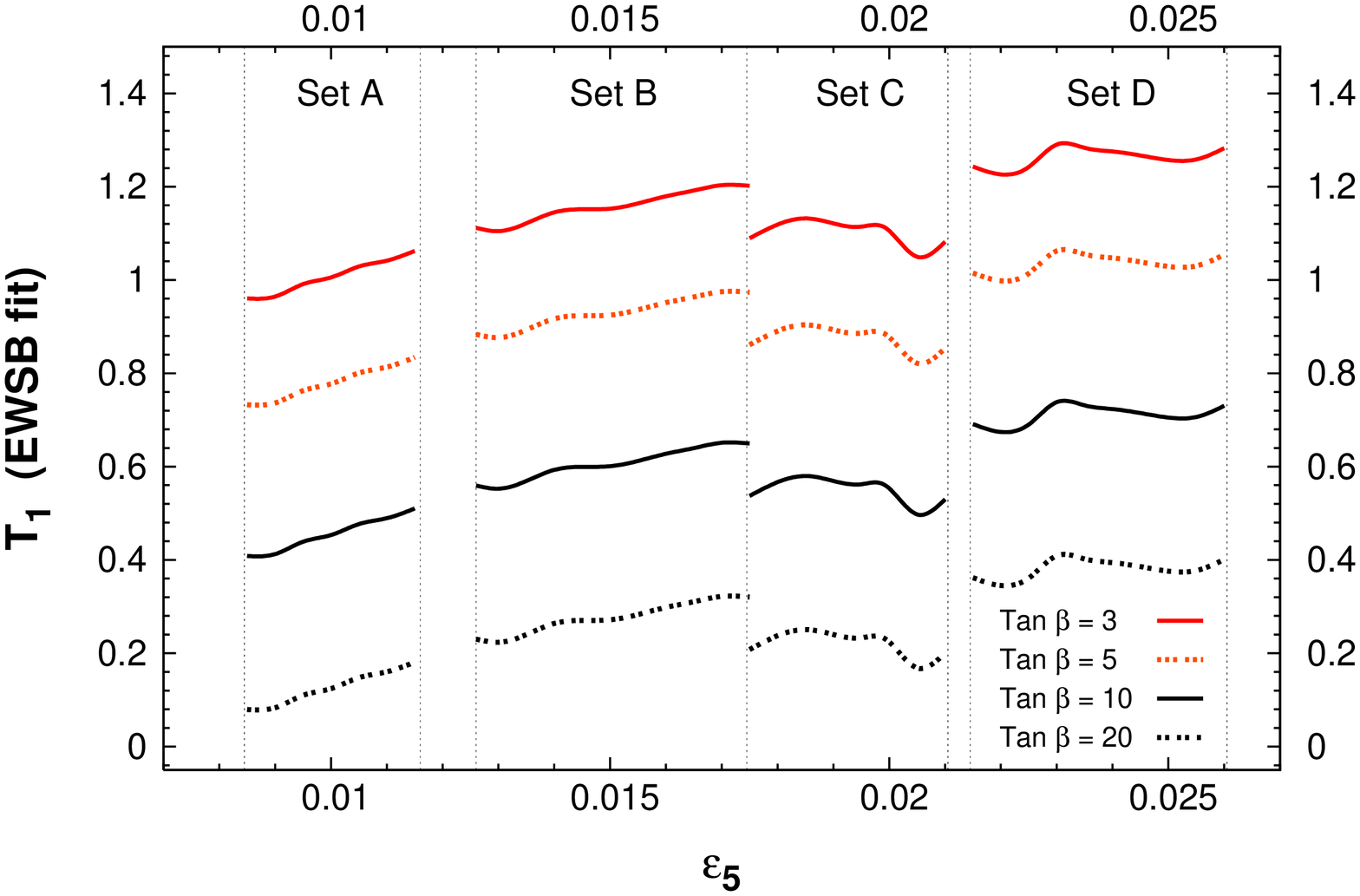,width=55mm,angle=270,clip=} &
	\psfig{file=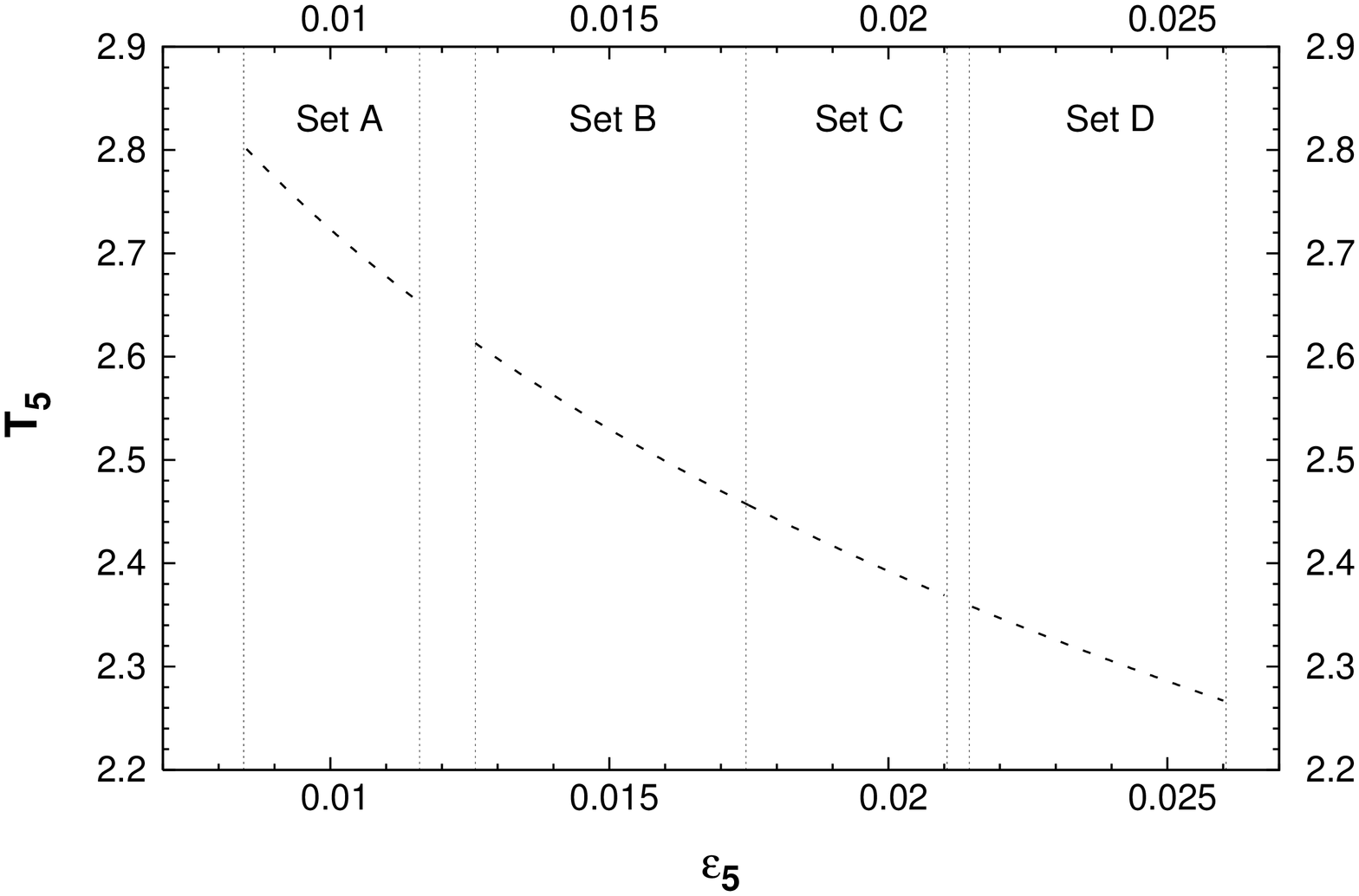,width=55mm,angle=270,clip=} 
    \end{tabular}
    \caption{Diagonal lattice moduli, $T_1$ (left) and $T_5$ (right), 
    as a function of $\varepsilon_5$~\cite{EMT-2_06}. For the case of
    $T_1$, we consider
    several values of $\tan \beta=$3, 5, 10 and 20. For $T_5$ the
    dotted line denotes the prediction of the orbifold.} 
    \label{fig:orbifold:T5:T1:e5}
  \end{center}
\end{figure}
\begin{figure}
  \begin{center} 
\psfig{file=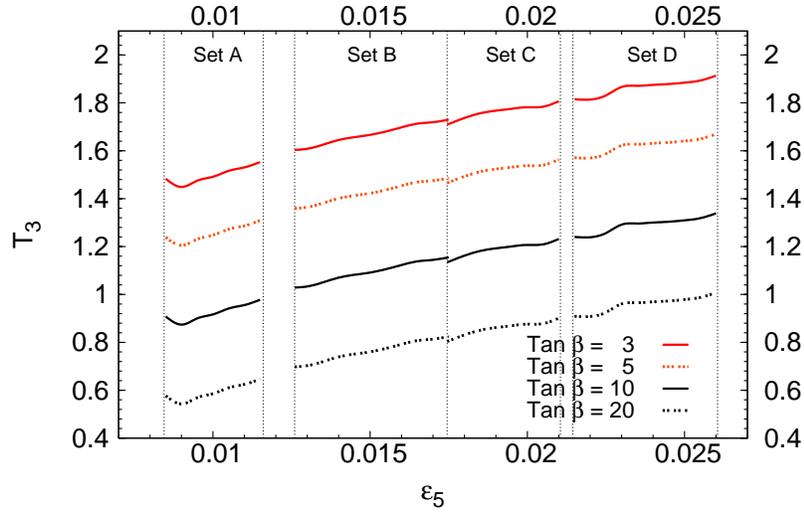,width=75mm,angle=270,clip=} 
    \caption{$T_3$-moduli dependence on $\varepsilon_5$, for $\tan
    \beta=3,5,10,20$.}  
    \label{fig:orbifold:T3:e5}
  \end{center}
\end{figure}
\begin{figure}
  \begin{center} \hspace*{-10mm}
    \begin{tabular}{cc}
	\psfig{file=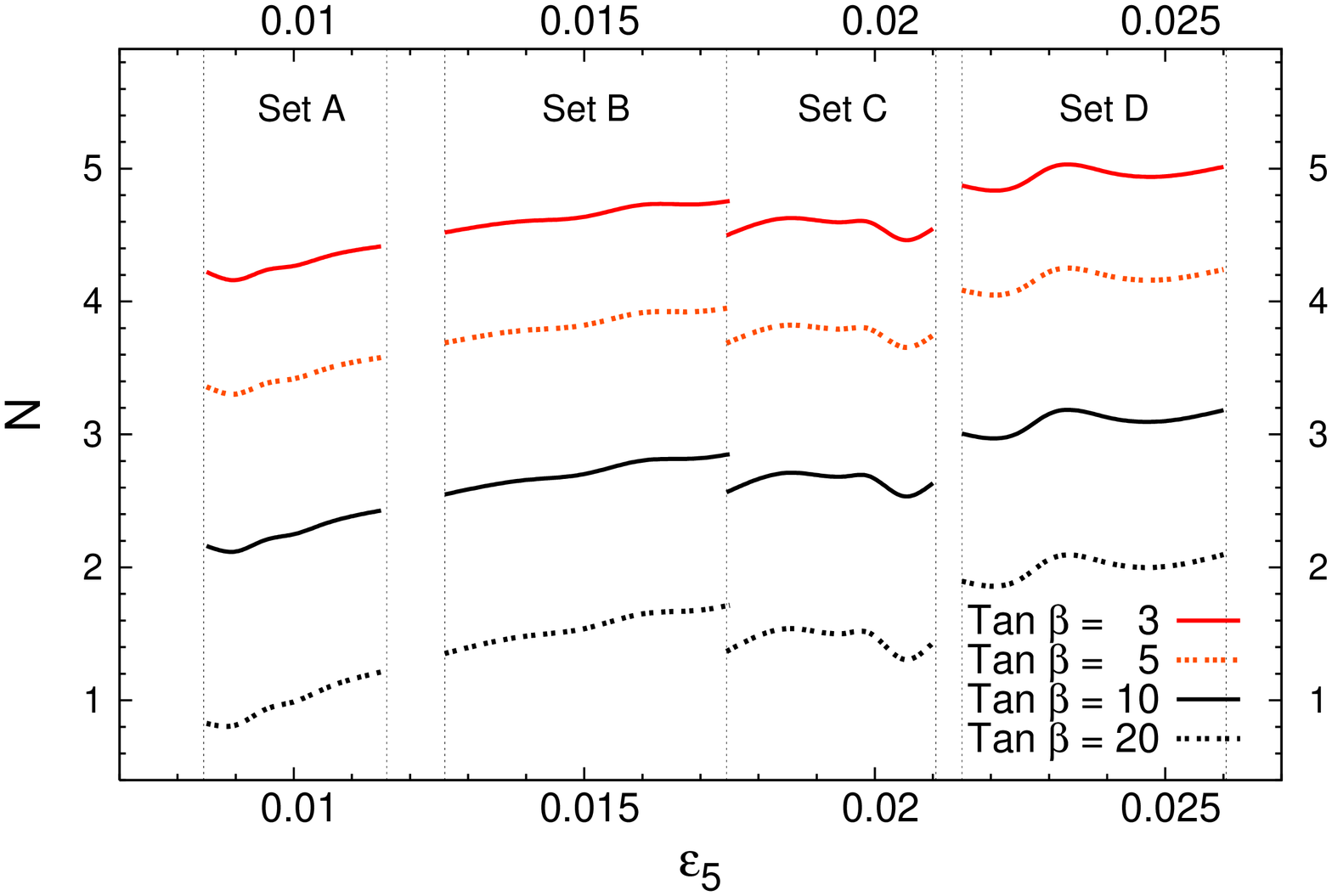,width=55mm,angle=270,clip=} &
	\psfig{file=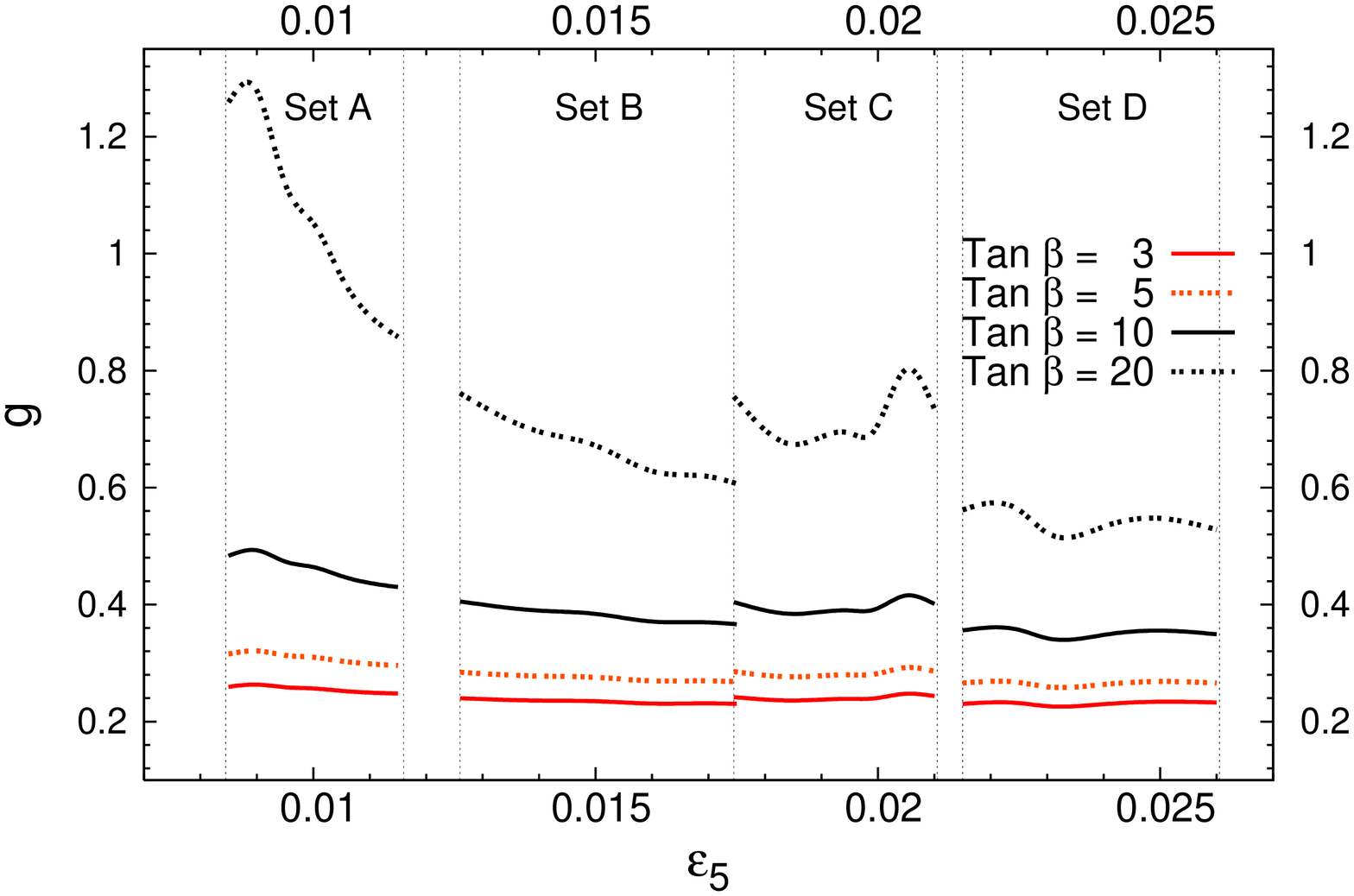,width=55mm,angle=270,clip=} 
    \end{tabular}
    \caption{Orbifold normalisation constant $N$ for
      $\tan{\b}=3,5,10,20$ as a function of $\varepsilon_5$ (left), and
      corresponding values of the heterotic gauge coupling constant
      $g$ (right).} 
    \label{fig:orbifold:Ng:e5}
  \end{center}
\end{figure}
where we have used the relation between the orbifold parameters given
by Eq.~(\ref{af:alpha:beta}). In Fig.~\ref{fig:orbifold:T5:T1:e5} we
display (as presented 
in Ref.~\cite{EMT-2_06}) the diagonal moduli 
$T_1$ and $T_5$ for different values of $\tan \beta$. In Figure
\ref{fig:orbifold:T3:e5} we 
show the correlation between those values of the moduli $T_3$
compatible with correct charged-lepton masses and  
$\varepsilon_5$, for different values of $\tan \beta$. 
From all plots we are led to verify
that $T_i\sim 1$, implying that the 
sizes of the radii in each orbifold sublattice are 
comparable and of order
\bea
R_i\,\approx \,\frac{4\,\pi}{3^{1/4}}\, \approx \,10\,.
\label{Rapprox}
\eea
As it was previously shown in Section \ref{sec_chlepmasses}, 
Eq.\,(\ref{Tfromeps}), the upper  
bound for the suppression factors $\varepsilon_i$ is 3, above which the
moduli are no longer a positive quantity. In the case of 
$T_1$ and $T_3$, the values are always fixed by the choice of a given
$\tan{\b}$, as we can see in the previous plots. This in turn implies
the existence of an upper limit for $\tan{\b}$, above which the
moduli become zero. In general, this occurs for values of $\tan{\beta}$ 
between 30 and 40, depending on the choice of the remaining orbifold
parameters. 

In the first analysis of Ref.~\cite{EMT-2_06}, the constraints on the
orbifold parameters derived from the quark sector had already allow to
hint towards a range for the product $gN$, $1.03 \lesssim g\,N
\lesssim 1.16$. The inclusion of the bounds arising from considering
the lepton sector finally allows to refine the knowledge of these
orbifold parameters. 
On the left hand-side of Fig.~\ref{fig:orbifold:Ng:e5} we show the value 
of the orbifold normalisation constant $N$ for different values of
$\tan{\b}$ as a function of $\varepsilon_5$. From this 
plot, using Eq.~(\ref{gNlimit}), we can derive the value of
the heterotic gauge coupling constant $g$. The result 
is presented on the right hand-side of Fig.~\ref{fig:orbifold:Ng:e5}. 
As we observe, for the chosen regimes of $\tan{\b}$, the value of  
$g$ varies between $\approx (0.2-1.2)$. 
It is worth noticing here that a value of $g$ of order 1, which is
compatible with the above result, was obtained in 
the orbifold models with three Higgs families analysed in~\cite{Munoz:2001yj},
in order to solve the discrepancy between the
unification scale predicted by the heterotic superstring 
($\approx g \, 5.27\times 10^{17}$ GeV) 
and the value deduced from LEP experiments 
($\approx 2\times 10^{16}$ GeV).

%
%
%
\section{Neutrinos}\label{sec_neutrinos}
As seen from the previous sections, after having imposed the
requirements of viable quark masses and mixings, as well as the
correct charged lepton masses, many of the orbifold parameters have
already been constrained. The question that remains to be answered is
whether or not the present neutrino data can be reproduced. 
In the following subsections, we will discuss how the orbifold model
allows us to deal with the problem of neutrino masses, providing 
several mechanisms
that can potentially account for an experimentally viable mass spectrum and
MNS matrix, including in some of the cases the generation of an effective
seesaw mechanism.

\subsection{Dirac neutrino masses without seesaw}
In the present orbifold model, the simplest way of obtaining massive neutrinos
is to assume that the latter are Dirac particles, and 
introduce a Yukawa term, coupling left- and right-handed
neutrinos to the up-type Higgs fields.
Accordingly, the Dirac mass matrix for the neutral leptons is given by:
\bea
{\cal
  M}^{\nu}\,=\,g\,N\varepsilon_1\,\varepsilon_3\,a^L\,a^{{\nu}^c}\, 
B^L \,A^u\, B^{\nu^c}\,
=\,g\,N\,\varepsilon_1\,\varepsilon_3\,a^L\,a^{{\nu}^c}\,
\left( \begin{array}{ccc}
v_2\, \varepsilon_5^2\, \beta^{L}\,\beta^{\nu^c} & 
v_6\,\varepsilon_5^2\,\beta^L  & v_4\,\epsilon_5\,\alpha^{\nu^c}\,\beta^L \\
v_6\,\varepsilon_5^2\,\beta^{\nu^c}  & v_4 & 
v_2\,\alpha^{\nu^c} \\
v_4\,\varepsilon_5\,\alpha^{L}\, \beta^{\nu^c}  & v_2\,\alpha^L  & 
v_6\,\alpha^{L}\,\alpha^{\nu^c}/\varepsilon_5^2
\end{array}\right)
\,,
\label{finalnudirac}
\eea
where $A^u$, $a^{L,\nu^c}$ and $B^{L,\nu^c}$ are defined in 
Eqs.~(\ref{AuAd:bfFI}) and (\ref{af:def}\,-\,\ref{alpha:beta:def}).
As shown in~\cite{Abel:2002ih}, unless some fine-tuning is introduced
in the model 
the use of these terms without the addition of Majorana couplings
gives rise to excessively heavy neutrinos.
This can be easily understood by noticing that all the parameters
involved in Eq.~(\ref{finalnudirac}) are completely determined from
the quark and charged-lepton sectors, the only exception being 
$\alpha^{\nu^c}$ (and thus $\beta^{\nu^c}$ and $a^{\nu^c}$). Thus, the
mass eigenvalues of the Dirac neutrinos can be approximately written as:
\be
m_{\nu_i}\,\approx \,\varepsilon_1\,\frac{a^{\nu^c}}{a^{e^c}} \,m_{l_i}
\,,
\label{numassbound}
\ee
leading to the relation
\be
\frac{m_{\nu_i}}{m_{l_i}}\,\approx\,
\varepsilon_1\,\frac{a^{\nu^c}}{a^{e^c}}\, \sim 10^{-7} \,.
\label{numassbound2}
\ee
Regarding the three parameters appearing in the previous equation,
$\varepsilon_1$ is defined by the chosen value of $\tan \beta$ 
(see~\cite{EMT-2_06}). For  
$\tan \beta$ between 3 and 20, values of
$\varepsilon_1$ compatible with realistic quark masses lie in the range
$\varepsilon_1\approx 0.2-2$. The factor $a^{e^c}$ has been determined
from the charged-lepton sector, $a^{e^c}\approx 0.1$. Thus the remaining
free parameter in Eq.~(\ref{numassbound2}) is $a^{\nu^c}$, 
which depends on $\alpha^{\nu^c}$ and
$\varepsilon_5$ in the following way (see
Eq.~(\ref{af:alpha:beta})): 
\be
a^{\nu^c}\,=\,\frac{(1-\alpha^{\nu^{c^2}})^{1/2}}{\alpha^{\nu^c}}
\,\frac{\varepsilon_5}{(1\,-\,\varepsilon_5)^{1/2}}\,.  
\label{anufromalfa}
\ee
Suppression of the light neutrino masses requires values of 
$\alpha^{\nu^c}$ very close to 1, forcing $a^{\nu^c}$ to be very small.
In turn, this would imply that there are additional fields entering the  
FI breaking, with a very distinct mass hierarchy (much lighter VEVs), 
giving rise to terms of the form
\be
a^{\nu^c}\,=\,\frac{\hat{c}_2^{\nu^c}}
{\sqrt{\,|\hat{c}_1^{\nu^c}|^2+|\hat{c}_2^{\nu^c}|^2\,}} \,\approx \, 
10^{-7}\,{-}\,10^{-6}\,. 
\label{anufromFI}
\ee
To clarify the latter statement, and 
as an example, let us consider the case in which the factors
$\varepsilon^{\prime(\nu^c)}$ and 
$\varepsilon^{\prime \prime(\nu^c)}$, defined in Eq.\,(\ref{hatc:def}), 
are taken to be $\varepsilon^{\prime(\nu^c)}=1$ and
$\varepsilon^{\prime\prime(\nu^c)}=\varepsilon_1\varepsilon_3\approx 0.01$. 
In this case, 
Eq.~(\ref{anufromFI}) may be rewritten as
\be
\frac{c_2^{\nu^c}}{c_1^{\nu^c}} \,\approx \,10^{-5}\,{-}\,10^{-4}\,.
\label{anuFIexample}
\ee
In order to fulfil the above condition, we are compelled to
modify the original hypothesis of assuming the VEVs $c^f_i$ 
to be of the order of the FI breaking scale, i.e.  
$10^{16\,\textrm{-}17}$ GeV.
One possibility of obtaining the desired hierarchy between
$c_1^{\nu^c}$ and $c_2^{\nu^c}$
is to invoke the existence of effective non-renormalisable
couplings of the form
\be
\frac{\vev{C^2}}{M_{\textrm{string}}^2}\,C^{\nu^c}_2\,\xi_1\,\xi_2\approx
\,10^{-4}\times\,C_2^{\nu^c}\,\xi_1\,\xi_2 \,,
\label{non_ren-couplings}
\ee
where $\xi_1,\,\xi_2$ denote two extra-matter fields which 
should later mix with the $\nu^c$ field \cite{Abel:2002ih}. Although this possibility
may solve the discrepancy between the FI-breaking scale and the one
needed to comply with realistic neutrinos, the introduction
 of non-renormalisable couplings sets an undesired arbitrariness 
in the mass scales used to generate the fermion masses. In this sense,
it seems preferable to find another way to generate
neutrino masses without the addition of higher-order operators.

Another possible solution to this problem could lie in the
assumption of a more involved mixing of the fields
participating in the FI breaking, as will be presented
in Section~\ref{FIseesaw}. 
Nevertheless, a more straightforward and simple possibility consists
of assuming that the neutrinos are Majorana particles. 
In this case, one allows the presence of Majorana terms in the superpotential, 
leading to a type-I seesaw mechanism. There are several possible ways
of implementing a seesaw mechanism in the context of these orbifold
models, and we pursue this topic in the following subsections. 
%
%
%
%
\subsection{Neutrino masses via a type-I seesaw}\label{sec:seesaw}
As first proposed in~\cite{Abel:2002ih}, the introduction of a seesaw
mechanism can be easily achieved by considering a Majorana term in the
superpotential, arising from the coupling of three extra scalars (of
the low-energy spectrum) as follows:
\bea
W^{\nu}\,\sim\,  H^u\, L\,\nu^c + S\,\nu^c\, \nu^c\ ,
\label{seesaw}
\eea
where $S$ are singlets assigned to the following fixed-point
components in the first two sublattices: 
\bea
S\ \,\,\,\, \times \ \times
\label{subl2}
\eea
Under this assumption, when the singlets develop a VEV, 
a Majorana mass for the right-handed neutrinos
is generated. In the seesaw limit, where the latter VEVs are much
heavier than the EW scale, the effective mass matrix for the light
neutrinos is then 
\bea
m_\nu^\text{eff} \, \approx \,
\mathcal{M}^{\nu }\, (\mathcal{M}^{\nu^c})^{-1}\,{\mathcal{M}^{\nu }}^{T}
\label{lightn}\,,
\eea
where $\mathcal{M}^{\nu }$ is given in Eq.~(\ref{finalnudirac})
and $\mathcal{M}^{\nu^c}$ arises from the coupling $S\,\nu^c\nu^c$, and is thus
defined as 
\bea
{\cal M}^{\nu^c}\,=\,g\,Na\,^{{\nu}^c}\,a^{{\nu}^c} \,B^{\nu^c} 
\,A^{s} \,B^{\nu^c}\,
\label{finalnuda}
\eea
with
\bea
A^{s}\,=\,\left( \begin{array}{ccc}
s_1 & s_3\varepsilon_5  & s_2\varepsilon_5 \\
s_3\varepsilon_5  & s_2 & s_1\varepsilon_5 \\
s_2\varepsilon_5  & s_1\varepsilon_5  & s_3
\end{array}\right)
\label{quarkmasses22}\,,
\eea
$s_{1,2,3}$ being the singlet VEVs. Note that in
Eq.~(\ref{lightn}) the mixing $B^{\nu^c}$ cancels,  
so that the only free parameters in the mass matrix will be the VEVs
$s_i$. It is also important to stress at this point that the Majorana
mass term in Eq.~(\ref{lightn}) is clearly non-diagonal, with a
structure which is determined from the orbifold (analogous to what occurs
for all the Dirac mass terms). 

The study of the parameter space generated by $s_i$ (for different 
regimes of the other parameters) reveals that it is possible  
to generate light neutrino masses of the desired order of magnitude, 
in good agreement with the experimentally measured mass squared 
differences between the three species, $\Delta m^2_{21}$ and
$|\Delta m^2_{31}|$ (see, for 
example,~\cite{GonzalezGarcia:2007ib}),  
\bea
\Delta m^2_{21} 
    & = & 7.9\,_{-0.28}^{+0.27}\,\left(_{-0.89}^{+1.1}\right)\,
\textrm{eV}^2 ,\label{delta_m_nu_theor}\\ 
\left|\Delta m^2_{31}\right|
 & = & 2.6 \pm 0.2\,(0.6) \times 10^{-3}\,
\textrm{eV}^2\,.
\label{delta_m_nu_theor2}
\eea
To illustrate this mechanism, let us define the seesaw
mass matrix as it would arise from the following point in the orbifold
parameter space, compatible with realistic quark (Set B,
Eq.~(\ref{setBquark})) and charged-lepton masses, 
\bea
\varepsilon_5=0.0126,\qquad \varepsilon_3=0.170, \qquad
\varepsilon_1=0.450 \,,\nonumber \\ 
\alpha^{u^c}=0.089,\qquad \alpha^{d^c}=0.295, \qquad \alpha^{L}=0.129\,.
\label{examplenuseesaw}
\eea
Setting $\tan{\beta}=5$, 
the only remaining parameters in the model are the singlet VEVs
$s_1,\,s_2,\,s_3$. 
The choice of the following values
\bea
\{s_1,\,s_2,\,s_3\}=\{2.45\times 10^{9},\,8.89\times
10^{12},\,1.32\times 10^{12}\}\ \textrm{ GeV}\,,
\label{singlet_values}
\eea
gives us a ``normal hierarchy''  light neutrino spectrum, 
\bea
\{m_{\nu_1},\,m_{\nu_2},\,m_{\nu_3}\}\,=\,\{5.39\times
10^{-8},\,9.13\times 10^{-3},\,5.65\times 10^{-2}\}\ \textrm{ eV},
\label{nu_mass_values}
\eea
leading to mass squared differences in good agreement with the 
experimental range of Eqs.~(\ref{delta_m_nu_theor},\,\ref{delta_m_nu_theor2}).
Even though this implementation of a type-I seesaw mechanism can lead
to a viable light neutrino spectrum, there are 
two drawbacks to this formulation. The first one comes
from the high scale required by the Majorana singlets ($10^{9-12}$ GeV). 
Again, a possible explanation of this high scale is to assume the
fields $S_i$ as effective 
non-renormalisable FI fields (analogous to the ones suggested in
Eq.~(\ref{non_ren-couplings}))  
or to allow a more complicated FI mixing which would translate into a
further suppression of the Yukawa couplings  
(see Section \ref{FIseesaw}, below).
The second shortcoming stems from a failure in reproducing the
observed mixing in the leptonic sector, as parameterised by the 
the $U_\text{MNS}$ matrix
\bea
U_{\textrm{MNS}}=
\left( \begin{array}{ccc}
c_{12}c_{13} & s_{12}c_{13} & \pm s_{13}\\
-s_{12}c_{23}\mp s_{23}s_{13}c_{12} & c_{12}c_{23}\mp 
s_{23}s_{13}s_{12} & s_{23}c_{13}\\
s_{12}s_{23}\mp s_{13}c_{23}c_{12} & -s_{123}c_{12}\mp 
s_{13}c_{23}c_{12} & c_{23}c_{13}
\end{array}\right)
\label{MNS_matrix_CP}\,,
\eea
where, for simplicity, we use the CP-conserving parametrisation. The
mixing angles $\theta_{12}$, $\theta_{23}$, $\theta_{13}$, are 
experimentally\footnote{We employ 
the values given in~\cite{GonzalezGarcia:2007ib}.}
given by
\bea
&\theta_{12} = & 33.7 \pm 1.3 \left(^{+4.3}_{-3.5} \right)\,, \\
&\theta_{23} = & 43.3\,^{+4.3}_{-3.8} \left(^{+9.8}_{-8.8} \right)\,, \\
&\theta_{13} = & 0\,^{+5.2}_{-0.0} \left(^{+11.5}_{-0.0} \right)\,,
\label{MNSangles-exp}
\eea
where the angles are expressed in degrees.
With the choice of orbifold parameters used in the previous example, 
Eq.~(\ref{examplenuseesaw}), we find that in this case the 
mixing angles in the $U_\text{MNS}$ matrix are
\bea
\{\theta_{12}^{\textrm{\,orbifold}},\,
\theta_{23}^{\textrm{\,orbifold}},\theta_{13}^{\textrm{\,orbifold}}\}\,=
\,\{48.4,\,54.5,\,17.9\,\}\,.
\label{MNSangles-orbifold}
\eea
As can be verified, the above values lie considerably above the ones allowed
by the experimental bounds. 
The associated $U_\text{MNS}$ would then be given by
\bea
U^{\textrm{orbifold}}_{\textrm{MNS}}=
 \left( \begin{array}{ccc}
 0.998 & 0.045 & 0.022\\
 0.050 & 0.87 & 0.48\\
 0.002 & 0.48 & 0.88
 \end{array}\right).
 \label{MNS_matrix_theor}
 \eea
This matrix contains nearly the desired mixing for the
second and third generations, but fails 
in reproducing the mixing for the first generation of neutrinos.
This behaviour is generic to the surveyed 
orbifold parameter space, where we have systematically found that no
more than two-generation mixing can be satisfied.
By varying the singlet VEVs other mixing possibilities 
can be achieved, but one generation of neutrinos never has a viable 
mixing with the other two. 
This appears to be a general feature of the present
orbifold model, in the sense that it is extremely difficult to
simultaneously accommodate the observed near-maximal mixing in the
lepton sector and the small one evidenced in the CKM matrix.

\vspace*{3mm}
A second possibility of implementing a type-I seesaw, without the need
of considering a more intricate FI breaking, consists in assuming the
existence of an intermediate scale.
In principle this scale is not predicted by the orbifold formulation, 
but it would nevertheless allow to accommodate the experimental 
data in view of 
orbifold-derived neutrino Yukawa couplings. 
In particular, in this case one is allowing for additional sources
of unconstrained mixing in the lepton sector, stemming from heavy
Majorana neutrino interactions.
Thus the effective light neutrino mass matrix is
obtained from the seesaw equation, and given by
\begin{align}\label{neut:mass}
& m_\nu\,=\,\mathcal{M}^{\nu }\, (M_{R})^{-1}\,{\mathcal{M}^{\nu }}^{T} \ , \\ 
& U^T_\text{MNS}\, m_\nu \, U_\text{MNS}\,=\,m_\nu^\text{diag}\,,
\end{align}
where $\mathcal{M}^{\nu }$ is defined
in Eq.~(\ref{finalnudirac}) and $M_R$ is  the Majorana mass matrix,
whose values are not determined by orbifold considerations. 
In general, $U_\text{MNS}$, $m_\nu^\text{diag}$ are known
and a very simple structure is adopted for $M_R$ (namely a diagonal
matrix) in order to derive the unknown Yukawa couplings.
In the present approach, we do know the Yukawa couplings (from the orbifold
construction, which at this stage has become strongly constrained),
and phenomenological viability of the orbifold scenario indirectly 
suggests the 
structure of $M_R$. Noticing that the seesaw equation can be rewritten as
\begin{equation}
\mathcal{M}^\nu\, U_\text{MNS}\,(m_\nu^\text{diag})^{-1} \,U^T_\text{MNS}\, 
{\mathcal{M}^\nu}^T\,=\, M_R\,,
\end{equation}
we obtain $M_R$ as required to comply 
with data on neutrino masses and mixings.
It is important to notice that we are not 
working in a basis where the charged lepton Yukawa couplings are diagonal, 
implying that the $U_\text{MNS}$ matrix is defined as 
\begin{equation}
U_\text{MNS}\,=\,{V^E_L}^\dagger\,U^n\,,
\end{equation}
where $V^E_L$ is the unitary matrix that rotates the left-handed
charged lepton fields, so to diagonalise $\mathcal{M}^e$, 
while $U^n$ is the matrix that diagonalises the symmetric 
neutrino mass matrix, $m_\nu$. 

\begin{figure}
  \begin{center} 
\psfig{file=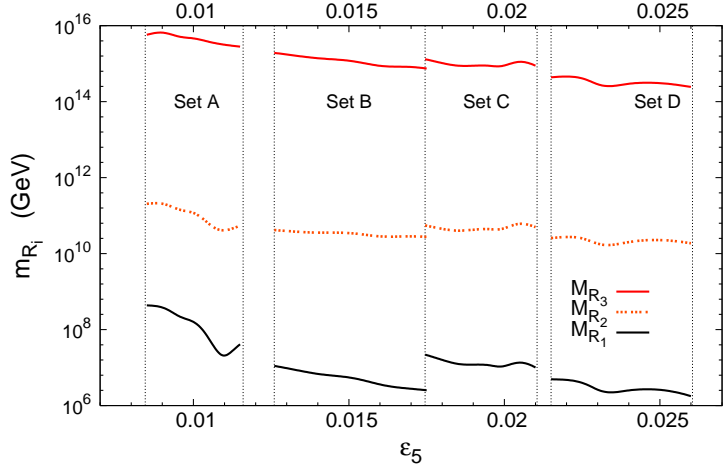,width=70mm,angle=270,clip=} 
    \caption{Eigenvalues of $M_R$ (in GeV) as a function of $\varepsilon_5$, 
      for the input orbifold parameter sets A, B, C and D. We take
      $\tan \beta=5$, $\theta_{13}=8^\circ$, and 
      $m_{\nu_i}=\{10^{-5}, 0.0089, 0.0509\}$ eV.} 
    \label{fig:orbifold:MR:e5}
  \end{center}
\end{figure}

In Fig.~\ref{fig:orbifold:MR:e5}, we depict the eigenvalues of $M_R$,
as a function of $\varepsilon_5$, for the input orbifold parameter
sets A, B, C and D. Leading to this figure we have assumed  $\tan
\beta=5$, $\theta_{13}=8^\circ$, a regime of 
$\alpha^{\nu^c} \approx \varepsilon_5$, and a normal hierarchy 
for the light neutrino spectrum, namely 
\begin{equation}
m_{\nu_i}\,=\,\{10^{-5}, 0.0089, 0.0509\}\,\mathrm{eV}\,. 
\label{lightnuNH}
\end{equation}
As can be seen from Fig.~\ref{fig:orbifold:MR:e5}, the orbifold
structure would indeed suggest the existence of heavy Majorana
neutrinos, whose masses would lie in the $10^{6}-10^{8}$ GeV and 
$10^{14}-10^{16}$ GeV ranges for the lightest and heaviest states,
respectively.
Naturally, the heavy spectrum strongly reflects the input
parameters, with the most important role being played by
$\alpha^{\nu^c}$, $\tan \beta$ and the hierarchy of the light
neutrinos.
Essentially $\tan \beta$ translates in an overall factor, and
having normal/inverted hierarchy or quasi-degenerate light neutrinos 
mostly affects the $m_{R_i}$ pattern. On the other hand, the chosen
$\alpha^{\nu^c}$ range can have a crucial impact: while values of 
$\alpha^{\nu^c} \approx \varepsilon_5$ (as used for
Fig.~\ref{fig:orbifold:MR:e5}) lead to  $m_{R_1}$ masses in the
$10^{6}-10^{8}$ GeV range, larger values, close to 1, can even give rise to
masses as small as $\mathcal{O}$(TeV). 
The phenomenological implications of the
latter regime would be extensive, and we do not address them
here.\\

Finally, and as an illustrative example, 
we present the complete $M_R$ matrix structure, for the orbifold
set of parameters taken in  
Eq.~(\ref{examplenuseesaw}), $\tan \beta=5$, $\theta_{13}=1^\circ$,
$\alpha^{\nu^c}=\varepsilon_5$ and 
the light-neutrino spectrum of Eq.\,(\ref{lightnuNH}):
\begin{equation}\label{MRnum}
M_R\,=\,\left(
\begin{array}{ccc}
3.86\times 10^{11} & 1.73\times 10^{12} & -3.35 \times 10^{13}\\
1.73\times 10^{12} & 7.73\times 10^{12} & -1.50\times 10^{14}\\
-3.35 \times 10^{13} & -1.50\times 10^{14} & 2.92\times 10^{15}
\end{array} \right)\, \text{GeV}\,.
\end{equation}
Using this matrix one can check that both 
a high seesaw scale and additional mixings involving the right-handed
neutrinos should be invoked in order to reproduce the  
correct neutrino masses and mixings. The eigenvalues for the Majorana
mass matrix (which correspond  
to the masses of the heavy neutrinos) are
\bea
\{m_{{R_1}},\,m_{{R_2}},\,m_{{R_3}}\}=
\{1.43\times 10^7,\,2.12\times 10^{10},\,2.93\times 10^{15}\}\,\textrm{ GeV}, 
\label{Right_nu_mass_values_prediction}
\eea
From these values one can see that, due to the mixing in the upper-left
$2\times 2$ block matrix in Eq.\,(\ref{MRnum}), 
we encounter a very suppressed mass eigenvalue. For different
regimes in the relevant parameters  
considered ($\tan{\b}$, $\theta^{\textrm{MNS}}_{13}$, $\alpha^{\nu^c}$
and the light-neutrino mass spectrum) 
one can check that this eigenvalue may be even sufficiently small to
lie at the TeV scale, being thus potentially  
detectable. The other eigenvalues, as can be seen in
Fig.~\ref{fig:orbifold:MR:e5}, remain always heavy, 
between $10^{10}$ and $10^{16}$ GeV. 

Although phenomenological viable, these last implementation of a
type-I seesaw is, as previously mentioned, neither related to the geometry
of the orbifold, nor to the dynamics associated with FI breaking. In
what follows, we pursue one final avenue, possibly leading to a more
appealing seesaw realisation.

%
\subsection{A viable seesaw from the FI breaking}\label{FIseesaw}
As mentioned, a third possibility for reproducing the observed
neutrino masses and mixings may be related with assuming a more
complex FI breaking. Here, we briefly outline the idea, for the
simplest case of one generation. 
Let us then assume that in addition to the $L$ and $\nu^c$ fields, 
which have the standard location 
\begin{equation} 
L\, \leftrightarrow \, \bullet \,\bullet \quad \quad \quad
\nu^c \, \leftrightarrow \, \times \times \,,
\end{equation}
there are additional matter fields (triplets, doublet or singlets)
$\zeta_i$, coupled to the $C_i$ fields which develop very large VEVs, 
thus inducing FI breaking. One can assume that
these fields have the following
assignments with respect to the first two sublattices:
\begin{align}
&
\zeta_1 \, \leftrightarrow \, \bullet \, \bullet \quad \quad 
\zeta_2 \, \leftrightarrow \, \times \, \circ  \quad \quad 
\zeta_3 \, \leftrightarrow \, \bullet \, \times \nonumber \\
&
C_1  \, \leftrightarrow \, \bullet \, \bullet \quad \quad
C_2  \, \leftrightarrow \, \circ  \, \times  \quad \quad 
C_3 \, \leftrightarrow \, \bullet \, \times
\end{align}
The latter $C_i$ develop VEVs, $c_i=\langle C_1 \rangle$, with 
$c_i \approx \mathcal{O}(10^{16-17})$ GeV. 
In principle, one can also have the following terms in the superpotential
\begin{equation}
C_1\,\zeta_1 \,L + C_2\,\zeta_2\, \zeta_1 +
C_2 \, \nu^c \,\zeta_3 + C_3\,\zeta_3 \,\zeta_3\,.
\end{equation}
With the above proposed lattice assignments, and after FI breaking,
this would lead to
\begin{equation}
c_1\,\zeta_1 \,L + 
\varepsilon_1 \, \varepsilon_3\,c_2\,\zeta_2\, \zeta_1 +
\varepsilon_1 \, c_2 \, \nu^c \,\zeta_3 + 
c_3\,\zeta_3 \,\zeta_3\,.
\end{equation}
In the basis defined by $(L\,\nu^c\,\zeta_1\,\zeta_2\,\zeta_3)^T$, one
would then arrive at the following ``mass matrix'' 
(again neglecting family dependence as a first approach),
\begin{equation}
{M^\nu}_{\text{FI}}\,=\,\left(
\begin{array}{ccccc}
0 & 0 & c_1 & 0 & 0\\
0 & 0 & 0 & 0 & \varepsilon_1 c_2 \\
c_1 & 0 & 0 & \varepsilon_1 \varepsilon_3 c_2 & 0 \\
0 & 0 & \varepsilon_1 \varepsilon_3 c_2 & 0 & 0 \\
0 & \varepsilon_1 c_2 & 0 & 0 & c_3
\end{array}\right)\,,
\end{equation}
with eigenvalues given by
\begin{equation}
m_i^\text{FI} \,= \, \left\{
0\,,\, \pm \sqrt{c_1^2\,+\, \varepsilon_1^2 \, \varepsilon_3^2
  \,c_2^2}\,,\, \frac{1}{2}\,\left(c_3 \, \pm \,\sqrt{c_3^2 \,+\, 
4\, \varepsilon_1^2\,c_2^2}\right) 
\right\}\,.
\end{equation}
Further assuming that we are in the limit where $\tan \beta$ is low
(favoured from several arguments, as discussed throughout this work),
$\varepsilon_1$ is smaller than unity. In this limit, and given that
$c_2 \sim c_3$, one would find
\begin{equation}
m_4 \,\approx \, c_3\,+\, \varepsilon_1^2\,c_3\,,
\quad 
m_5 \,\approx \, -\varepsilon_1^2\,c_3\,,
\end{equation}
thus implying the presence of a term in the superpotential behaving like
\begin{equation}
-\varepsilon_1^2\,c_3\,\nu^{c \prime}\,\nu^{c \prime}\,,
\label{majo:fi}
\end{equation}
with 
\begin{equation}
\nu^{c \prime}\,\propto \,\left( -c_3 \, + \,\sqrt{c_3^2 \,+\, 
4\, \varepsilon_1^2\,c_2^2}\right)\, \nu^c \, + 
\left(2\,\varepsilon_1\,c_2\right)\, \zeta_3.
\end{equation}
In the superpotential involving the MSSM fields, 
the term in Eq.~(\ref{majo:fi})
would effectively generate a Majorana mass term for the neutrino field.
Thus, an intermediate Majorana scale (lower than the FI breaking
scale, and much heavier than the EW scale) would naturally appear,
induced from the dynamics of FI breaking. Other mixings between 
$\zeta_i$ and the remaining matter fields could in principle occur, but
can be suppressed by some appropriate symmetry.

\vspace*{3mm}
Whether or not such a FI-seesaw would indeed reproduce the correct three
family neutrino masses and mixings is a question worth discussion. 
Additionally, one should also recall that the existence of heavy
Majorana neutrinos, with possibly complex Yukawa couplings (for a
discussion of how to implement CP violation in this class of orbifold
constructions, see Ref.~\cite{EMT-2_06}), offers the possibility of
generating the observed baryon asymmetry of the Universe from thermal
leptogenesis~\cite{Fukugita:1986hr}. This can be an involved question, 
given that as
seen in Section~\ref{sec:seesaw}, 
the scale of the lightest right-handed neutrino can
range over several orders of magnitude. These two issues, and others
like the collider signatures of potentially light Majorana neutrinos
will be the subject of a subsequent analysis.

%
\section{Conclusions}\label{concs}

In this study we have aimed at completing the analysis of the
phenomenological viability of abelian $Z_3$ orbifold compactifications
with two Wilson lines. 
This class of models, which naturally includes three families 
of fermions and Higgs fields, offers the
possibility of obtaining realistic fermion masses and mixings,
entirely at the renormalisable level. 
The Yukawa couplings arise from the geometrical configuration of the 
orbifold, and since they are explicitly calculable, one obtains 
a solution to the flavour problem of the SM and MSSM.

Successfully reproducing the observed pattern of quark masses and mixings
already severely constrains the orbifold parameters. Furthermore, 
the presence of six Higgs doublets poses potential problems 
regarding tree-level FCNCs, which can nevertheless be avoided with a fairly 
light Higgs boson spectrum. 

Here we have addressed in detail the implications of this class of 
orbifold compactifications for the lepton sector. Regarding the charged 
leptons, we verified that the still unconstrained orbifold parameters
could easily account for the observed spectrum. Moreover, and even though
one is equally likely to encounter tree-level contributions to three-body 
LFV decays, the typical choices of Higgs soft-breaking masses (taken as
to comply with the bounds on neutral meson FCNC) ensure that the
predicted BRs lie several orders of magnitude below the experimental
bounds.

Regarding the neutrino sector, the orbifold model offers multiple
possibilities. Albeit promising, we verified that the hypothesis of
strictly Dirac neutrinos requires that the fields entering the FI
breaking should have extremely hierarchical VEVs, forcing to call upon
effective non-renormalisable couplings.
Implementing a type-I seesaw mechanism via
extra singlet fields whose interactions are dictated by the orbifold
configuration reveals to be equally difficult.
Complying with the measured mass squared differences favours 
VEVs for the Majorana singlets far higher than for the other fields. 
This again introduced the need to interpret these fields as
effective non-renormalisable fields. Additionally, this mechanism
fails in accommodating the current bounds on the neutrino mixing angles.

The need of additional mixing involving the Majorana singlet sector, and of 
an intermediate scale of about $10^{9-10}$ GeV motivated us to
consider a third possibility. We have thus assumed that the smallness of
the light neutrino masses is indeed explained by a type-I seesaw
mechanism, where nor the scale, nor the mixings of the heavy singlets
are predicted by the orbifold. In this case we verified that neutrino
masses and mixings can be easily obtained, with a particularly
interesting possibility which is that of a TeV-mass Majorana singlet.

In spite of the latter, it would be theoretically more appealing and
consistent to have neutrino masses and mixings strictly from geometrical
argumentations and/or from FI breaking. We pursued this challenging
possibility, finding that in the simplest one-generation case, 
a slightly more involved FI breaking can indeed give rise to a
Majorana mass term, with a scale far lower than that of FI breaking,
and much higher than the EW scale.

This final possibility is definitely worth further investigation. In
addition, one can also investigate the viability of generating the
observed baryon asymmetry of the Universe from thermal leptogenesis.
Having Majorana singlets that can be (although not necessarily) as
light as the EW scale, also poses interesting scenarios regarding
collider signatures. 
%
%
%
\section{Acknowledgements}
N.~Escudero is deeply grateful to the members of the Laboratoire de
Physique Th\'eorique, Universit\'e de Paris\,-Sud XI, 
for their kind hospitality in Orsay during the final stages of this
work. He also thanks E.~Arganda
for useful discussions. 

The work of N. Escudero is supported by the ``Consejer\'{\i}a de
Educaci\'on de la Comunidad de
Madrid - FPI Program'' and ``European Social Fund''.
The work of C. Mu\~noz is supported 
in part by the Spanish DGI of the
MEC 
under Proyectos Nacionales FPA2006-05423 and FPA2006-01105,
by the European Union under the RTN program 
MRTN-CT-2004-503369, and by the Comunidad de Madrid under Proyecto
HEPHACOS, Ayudas de I+D S-0505/ESP-0346. 
The work of A.~M.~Teixeira is supported by the French ANR
project PHYS@COL\&COS.

%
%
%

%
\end{document}